\PassOptionsToPackage{unicode}{hyperref}
\PassOptionsToPackage{hyphens}{url}
\documentclass[a4paper,10pt]{article}
\usepackage{xcolor}
\usepackage[margin=1in]{geometry}
\usepackage{amsmath,amssymb}
\usepackage{wrapfig}
\usepackage{floatrow}
\usepackage[font=it]{caption}
\usepackage{iftex}
\ifPDFTeX
  \usepackage[T1]{fontenc}
  \usepackage[utf8]{inputenc}
  \usepackage{textcomp} 
\else 
  \usepackage{unicode-math} 
  \defaultfontfeatures{Scale=MatchLowercase}
  \defaultfontfeatures[\rmfamily]{Ligatures=TeX,Scale=1}
\fi
\usepackage{lmodern}
\ifPDFTeX\else
\fi
\IfFileExists{upquote.sty}{\usepackage{upquote}}{}
\IfFileExists{microtype.sty}{
  \usepackage[]{microtype}
  \UseMicrotypeSet[protrusion]{basicmath} 
}{}
\makeatletter
\@ifundefined{KOMAClassName}{
  \IfFileExists{parskip.sty}{%
    \usepackage{parskip}
  }{
    \setlength{\parindent}{0pt}
    \setlength{\parskip}{6pt plus 2pt minus 1pt}}
}{
  \KOMAoptions{parskip=half}}
\makeatother
\usepackage{longtable,booktabs,array}
\usepackage{calc} 
\usepackage{etoolbox}
\makeatletter
\patchcmd\longtable{\par}{\if@noskipsec\mbox{}\fi\par}{}{}
\makeatother
\IfFileExists{footnotehyper.sty}{\usepackage{footnotehyper}}{\usepackage{footnote}}
\makesavenoteenv{longtable}
\usepackage{graphicx}
\makeatletter
\newsavebox\pandoc@box
\newcommand*\pandocbounded[1]{
  \sbox\pandoc@box{#1}%
  \Gscale@div\@tempa{\textheight}{\dimexpr\ht\pandoc@box+\dp\pandoc@box\relax}%
  \Gscale@div\@tempb{\linewidth}{\wd\pandoc@box}%
  \ifdim\@tempb\p@<\@tempa\p@\let\@tempa\@tempb\fi
  \ifdim\@tempa\p@<\p@\scalebox{\@tempa}{\usebox\pandoc@box}%
  \else\usebox{\pandoc@box}%
  \fi%
}
\def\fps@figure{htbp}
\makeatother
\providecommand{\tightlist}{%
  \setlength{\itemsep}{0pt}\setlength{\parskip}{0pt}}
\usepackage{bookmark}
\IfFileExists{xurl.sty}{\usepackage{xurl}}{} 
\usepackage{tcolorbox}
\tcbuselibrary{breakable,skins}
\usepackage{apacite}
\tcbset{colback=white}
\newcommand{\modelcardsep}{%
  \par\vspace{10pt}%
  \noindent\textcolor{gray!40}{\rule{\linewidth}{0.4pt}}%
  \par\vspace{8pt}%
}
\newenvironment{modelcardpanel}{%
  \modelcardsep%
}{%
  \par%
}
\renewenvironment{abstract}{%
  \begin{tcolorbox}[%
    enhanced,
    sharp corners,
    colback=gray!4,
    colframe=gray!40,
    boxrule=0.7pt,
    title=\textbf{Executive Summary},
    fonttitle=\large\bfseries,
    left=10pt,
    right=10pt,
    top=10pt,
    bottom=10pt,
    breakable]
    \setlength{\parskip}{6pt}
    \noindent
}{%
  \end{tcolorbox}
}
\hypersetup{
  pdftitle={Institutional AI Sovereignty Through Gateway Architecture: Implementation Report from Fontys ICT},
  pdfauthor={Ruud Huijts and Koen Suilen},
  hidelinks,
  pdfcreator={LaTeX via pandoc}}

\title{Institutional AI Sovereignty Through Gateway Architecture:
Implementation Report from Fontys ICT}
\author{Ruud Huijts\\\texttt{r.huijts@fontys.nl} \and Koen Suilen\\\texttt{k.suilen@fontys.nl}}
\date{December 2025}

\begin{document}
\maketitle

\vfill
\begin{center}
\small This whitepaper was developed through iterative collaboration with Claude Sonnet 4.5 (Anthropic). The AI served as an intellectual sparring partner in developing arguments, structuring content, and refining language. All factual claims were independently verified, and the authors retain full responsibility for the work's content and conclusions.
\end{center}
\vspace*{2em}
\clearpage

\begin{abstract}

To address the fragmented, high-risk and inequitable adoption of commercial AI tools, we designed and operated an institutional AI platform in a six-month, 300-user pilot. The aim was to demonstrate that a university of applied sciences can offer advanced AI capabilities on its own terms, with fair access for all students and staff, transparent risks, controlled costs and full alignment with European law.

Commercial AI subscriptions proved incompatible with that mission. They create unequal access, because only those who can pay receive robust AI support, and they introduce serious compliance risks through opaque data processing, non-EU hosting and provider-centric terms of use. At the same time, simply banning such tools is neither realistic nor desirable in education. The institution needs a way to make powerful AI available in a form that is sovereign, accountable and compatible with its public role.

Our solution is a governed gateway platform with three tightly integrated layers:

\textbf{Frontend}: A familiar, ChatGPT-style interface tied to institutional identity (single sign-on). It gives every student and staff member the same secure entry point to AI, makes model choice explicit and surfaces key information such as hosting region and indicative cost, turning each interaction into a small lesson in AI capabilities, risks and trade-offs.

\textbf{Gateway}: The operational core that translates institutional policy into technical enforcement. The gateway controls who may use which models, under what conditions and with what budget. It steers traffic to EU infrastructure by default, logs usage for audit and cost allocation and ensures that any non-EU processing is a conscious, documented exception rather than an accident.

\textbf{Provider Layer}: A catalog of commercial and open-source models. Each model is wrapped in an institutional model card and consent flow that spell out capabilities, limitations, costs, hosting location and risk profile. This consolidates scattered vendor documentation into a single governance interface and provides teaching material for AI literacy, privacy awareness and cost-conscious use.

The pilot showed that this layered approach works technically and organisationally. The platform ran reliably, with no privacy incidents and strong user adoption. It enabled the institution to steer traffic towards EU infrastructure by default, to manage usage and spending across groups and projects and to make model choices transparent to users. At the same time, it highlighted that neither commercial licensing, consortium purchasing nor full self-hosting offer a sustainable alternative: only the gateway pattern combines model diversity and rapid innovation with institutional control over access, data flows and documentation.

The central insight from the pilot is that technology alone is not enough. AI is no longer a support function to be aligned with strategy—AI is strategy itself. This shift demands dedicated leadership. Sustainable operation of such a platform depends on explicit governance: someone must decide which models are admissible under which conditions, oversee legal and ethical risks, steer budgets and ensure that the platform's governance features are actually used in teaching and assessment. Existing roles cannot absorb these responsibilities because AI's dynamic, evolving nature creates governance challenges that transcend traditional functional boundaries.

Based on our experience, we therefore recommend establishing a formal AI Officer function, implemented as an individual or a coordinated team. This role must combine technical literacy to assess models and logs, governance authority to set and enforce policy and educational responsibility to connect the platform to curricula and examinations. Without this dedicated function, AI-related decisions will remain ad-hoc and the institution's exposure to legal, ethical and strategic risks will grow. With it, the pilot shows that institutional AI sovereignty is not only possible but practical: a higher-education institution can operate its own powerful, multi-provider AI platform—provided it takes governing AI as seriously as it takes teaching it.

\end{abstract}

\setcounter{tocdepth}{1}
\newpage
\tableofcontents

\clearpage
\section{Part I: Why We Built This}\label{part-i-why-we-built-this}

\subsubsection{The Fontys ICT Context}\label{the-fontys-ict-context}

Fontys ICT is a university of applied sciences in the Netherlands where students, researchers, and teaching staff work on real-world challenges in the domain of ICT. The institution offers education at Associate Degree, Bachelor's, and Master's levels. Fontys ICT's vision on education is that students shape the professional field—a philosophy that demands infrastructure reflecting professional standards rather than consumer convenience.

We experienced the familiar patchwork of tools that followed the rise of generative AI. Students and faculty mixed ChatGPT, Claude, Gemini, and countless niche services based on personal preference; only Microsoft Copilot was institutionally licensed. The pace of new releases magnified the sprawl, and the institution felt compelled to respond.

Financial inequity became visible quickly. Faculty expensed premium subscriptions on personal cards, administrative staff processed reimbursements case by case, and students who could pay for ChatGPT Plus or Claude Pro outpaced peers limited to free tiers. Copilot, the lone sanctioned option, was judged weaker than the commercial tools students already knew.

Technical usage exposed bigger risks. Students wiring APIs into coursework chose providers for their generous free tiers, not their privacy posture. The result: fragmented contracts, unclear data-processing locations, unknown retention windows, and no way for faculty to verify which tools touched submitted work. Institutional risk became the sum of individual choices.

The turning point came in December 2024 when SURF, the Dutch research and education network, advised against deploying Microsoft 365 Copilot. Its DPIA flagged opaque data collection, incomplete access responses, fabricated personal data, and excessive telemetry retention.

That warning clashed with reality: Copilot sat inside our existing Office 365 licenses and Microsoft pitched it as the obvious next step. The contradiction forced a deeper question: if we cannot trust the vendor default, who decides which AI tools we do trust, on what criteria, and through what process?

SURF reduced the risk rating to “medium” in September 2025, but the underlying issue stood. Institutions need their own framework for evaluating AI tools against their values instead of inheriting the vendor’s priorities by default.

This question reveals the fundamental inadequacy of commercial subscription models for educational institutions. Educational institutions prioritize institutional values over commercial convenience, have compliance obligations extending beyond consumer protection, and face pedagogical requirements that commercial interfaces cannot satisfy. Commercial AI subscriptions optimize for individual consumer convenience and maximize provider revenue, while educational institutions require infrastructure aligned with institutional values, regulatory compliance frameworks, and pedagogical objectives that may directly conflict with commercial optimization targets.

We distilled the institutional requirements commercial subscriptions
failed to meet:

\begin{itemize}
\item \textbf{Educational equity}: Access cannot depend on personal
budgets. Premium subscriptions gave paying students obvious advantages in
coursework and experimentation while free-tier limits held others back.
\item \textbf{Budget management and cost control}: When faculty and staff
purchase individual commercial subscriptions—some reimbursed by the
institution, others not—the institution loses all visibility into total AI
spending, cannot track return on investment, and creates administrative
overhead processing scattered reimbursement requests. Beyond tracking,
commercial subscriptions at scale are economically unsustainable: providing
every staff member with a ChatGPT Plus subscription alone would cost
significantly more than operating a shared gateway infrastructure while
locking the institution into a single vendor's ecosystem.
\item \textbf{GDPR and AI Act alignment}: Deployers must provide clear
documentation of capabilities, limitations, and oversight
\cite{gdpr2016,aiact2024}. Vendors scatter that information across
policies and portals, making institutional-scale compliance impossible
when every user registers independently.
\item \textbf{Pedagogical control}: ICT students need programmable
infrastructure---API access, cost-performance analysis, fallback
strategies, and exposure to multiple architectures. Consumer chat
interfaces hide the complexity students must master.
\item \textbf{Research flexibility}: Professorships require tailored model
combinations, budget profiles that match research intensity, and APIs
that slot into existing development workflows. Consumer products do not
offer that configurability.
\end{itemize}

Consumer APIs did not solve the problem; they inherit the same opaque
data routing, depend on provider assurances, and tie research output to
personal accounts that vanish when students graduate. Without
institutional guardrails, operational and compliance risk remain with the
individual user.

These requirements converged into a fundamental institutional question:
how can we provide the full diversity of AI models and
capabilities our students and researchers require while maintaining
privacy protection, regulatory compliance, cost sustainability, and
alignment with our educational mission?

\subsubsection{Evaluating Alternative
Approaches}\label{evaluating-alternative-approaches}

Before committing to institutional infrastructure, we assessed
strategies commonly used across higher education, evaluating each against
equitable access, data protection with geographic transparency, and
pedagogical control.

\begin{enumerate}
  \item \textit{Commercial subscription consolidation.}

Institution-wide licensing with one or more vendors would simplify
access and user experience. However, the cost structure proved
prohibitive at institutional scale—universal staff subscriptions alone
would exceed gateway operating costs while delivering only a single
vendor's capabilities. Beyond economics, it constrained pedagogical
flexibility to provider roadmaps, left data-routing transparency
dependent on vendor assurances, and kept long-term pricing and model
availability outside institutional control.

  \item \textit{Consortium procurement (e.g., via SURF).}

Organizations like SURF negotiate bulk deals with AI providers on behalf 
of multiple institutions, securing better prices through collective 
purchasing power.~ However, SURF's DPIA on Microsoft 365 Copilot advised 
caution due to unresolved privacy risks, demonstrating that lower prices 
do not resolve data-protection concerns.~ Consortium procurement also 
introduces an extra management tier between institutions and providers, 
reducing flexibility and innovation capacity. While EduGenAI (currently 
under construction) might become part of future solutions, every AI 
decision outsourced to a provider limits learning opportunities for 
  students and staff. Consortium purchasing improves 
pricing but inherits core limitations for privacy-sensitive and 
  pedagogy-specific use cases.\footnote{This is our institutional assessment, 
  not a SURF position.}

  \item \textit{Self-hosting open-source models.}

Operating models on institutional hardware maximizes sovereignty and
avoids vendor lock-in. In our context, the required GPU capacity,
staffing, and lifecycle management would have diverted resources from
teaching and research, and would limit exposure to frontier models that
students encounter in professional practice.

  \item \textit{Gateway architecture (selected).}

A multi-provider gateway allowed us to preserve model diversity
while enforcing institutional policy on access, budgeting, and
EU-first routing. The gateway decouples pedagogy from vendor roadmaps
and turns compliance into configuration, rather than documentation
  alone. Importantly, the gateway architecture can incorporate elements of 
  alternatives 1--3 as needed: we can negotiate vendor subscriptions 
  (alternative 1) through the gateway, integrate consortium offerings like 
  EduGenAI (alternative 2) as routing options, or operate self-hosted models 
  (alternative 3) as internal endpoints. Rather than locking into any single 
  approach, the gateway provides institutional control over which strategies 
  to deploy and how to combine them. It also enables us to integrate evolving 
  external offerings—such as SURF's EduGenAI API (currently under construction)—
  without sacrificing institutional control. We can incorporate EduGenAI as one 
  provider among many, rather than replacing our infrastructure entirely and 
  losing the ability to set our own access policies, routing rules, and pedagogical 
configurations. Where external risk posture changes over time, the gateway 
can adapt providers and routes accordingly.
\end{enumerate}

\textit{Conclusion.}

The gateway approach provided the workable balance we needed: equitable,
transparent access with enforceable controls, and the agility to respond to
changing provider risk assessments and sector guidance.

\begin{table}[htbp]
\centering
\resizebox{\textwidth}{!}{%
\begin{tabular}[]{@{}lllll@{}}
\toprule\noalign{}
\textbf{Approach} & \textbf{Equity of Access} & \textbf{Data Sovereignty
\& Compliance} & \textbf{Pedagogical Flexibility} &
\textbf{Feasibility} \\
\midrule\noalign{}
Commercial Licensing & Medium & Low & Low & High \\
Consortium Procurement (e.g., SURF) & Medium-High & Medium & Low-Medium
& High \\
Self-Hosting & High & High & High & Low \\
Gateway Architecture (Selected) & High & High & High & Medium-High \\
\bottomrule\noalign{}
\end{tabular}%
}
\caption{Comparison of AI deployment approaches for educational institutions}
\label{tab:approaches}
\end{table}

\begin{center}\rule{0.5\linewidth}{0.5pt}\end{center}

\section{Part II: What We Built}\label{part-ii-what-we-built}

\subsubsection{Architecture Overview: Three-Layer Design
Rationale}\label{architecture-overview-three-layer-design-rationale}

We designed the system around clean separation of concerns across three
distinct layers, each addressing specific institutional requirements
while maintaining interfaces enabling independent evolution.

The frontend layer handles user interaction and identity integration. It
uses familiar interfaces to minimise adoption barriers while surfacing
model characteristics such as geographic processing locations and
expected use cases. Single Sign On through Active Directory keeps access
aligned with course and research groups, and the UI evolves in response
to teaching feedback.

The gateway layer translates institutional decisions into enforcement,
converting rules about availability, budgets, access scopes, and hosting
into technical constraints while abstracting provider-specific details.
Configuration updates follow governance changes as policies and
compliance frameworks develop.

The external provider layer supplies the model catalog. By maintaining
direct relationships with multiple vendors we preserve contractual
controls over data handling and geography, while the gateway abstraction
lets us add or retire providers as offerings, pricing, or risk profiles
change—without touching the frontend.

\begin{figure}[htbp]
\centering
\includegraphics[width=0.4\textwidth,keepaspectratio]{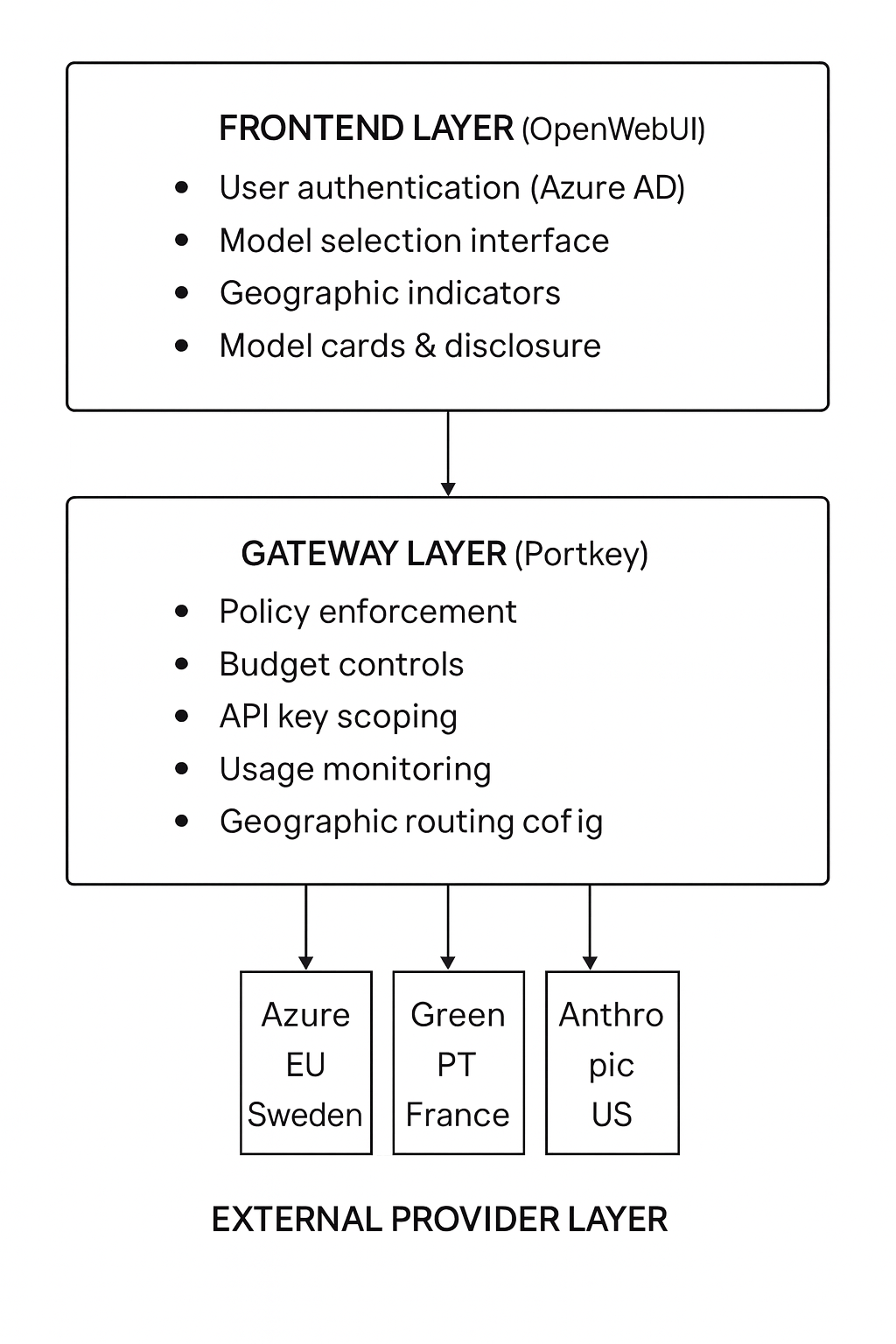}
\caption{Three-layer gateway architecture diagram showing frontend, gateway, and provider layers}
\label{fig:architecture}
\end{figure}

\subsubsection{Frontend Layer: OpenWebUI for Familiar User
Experience}\label{frontend-layer-openwebui-for-familiar-user-experience}

For the initial pilot we selected OpenWebUI (after evaluating LibreChat
and custom-built interfaces) because it integrates cleanly with our
gateway, supports straightforward model administration, benefits from an
active community, and matches the ChatGPT-style experience students and
staff already know. That familiarity avoids training overhead and keeps
attention on the differences we need to emphasise: transparent model
choice, hosting visibility, and access to diverse model families.

OpenWebUI implements several critical capabilities supporting our
institutional requirements:

Azure Active Directory authentication integration ensures that platform
access flows through existing institutional identity management.
Students and faculty authenticate using their institutional
credentials, eliminating separate account creation while providing
centralized audit trails, immediate access revocation upon departure,
and security standards consistent with other institutional services.

AAD integration also enables group-based access controls with
exceptional granularity aligned to organizational hierarchies and
course structures. Applied GenAI program students automatically receive
baseline model access through enrollment groups; research teams obtain
specialized models via professorship groups; administrative staff access
operational models through departmental groups. For particularly
sensitive scenarios, we can scale down to single-user, single-model
access with dedicated budget allocation---enabling high-intensity use
of expensive models without impacting shared institutional resources.

Explicit model selection is central to the pedagogy. Rather than default
routing, every chat session begins with users choosing a model from the set
their group permissions allow. The list is curated by governance policy so
students only see options that satisfy institutional requirements for data
sovereignty, risk, and budget. Each entry surfaces the trade-offs we want
them to reason about: hosting region is flagged (EU locales such as Sweden,
Germany, Netherlands versus US datacenters), relative cost appears as euro
icons benchmarked against a baseline, capability tags
highlight intended use cases, and compliance badges indicate whether extra
approval is required. A recommended default is preselected, but switching
models is a conscious action that triggers a tooltip summarising latency,
token limits, and privacy caveats. A prominent link opens the full model
card with technical, privacy, compliance, cost, and alternative references
(detailed in Section II.D). The workflow stays light, but every choice
exposes privacy, capability, and spend implications with enough context to
support informed decision-making.

Together, these interface elements transform model selection from
invisible background implementation detail into explicit decision-making
opportunity aligned with ICT educational objectives. Students learn to
evaluate technical trade-offs, consider privacy implications, assess
cost-performance relationships, and select appropriate tools for
specific contexts. Rather than AI remaining a black box that magically
produces outputs, students engage with the full complexity of
professional AI system integration.

\subsubsection{Gateway Layer: Portkey for Multi-Provider
Coordination}\label{gateway-layer-portkey-for-multi-provider-coordination}

The gateway layer turns governance into code. We chose Portkey (over
custom and other commercial gateways) because it unifies provider
integrations, issues scoped API keys with budget limits, enforces spend
controls at model/user/project levels, and records usage without logging
content—so decisions about availability, budgets, hosting, and monitoring
are enforced automatically rather than left to user discipline.

Multi-provider integration enables connection to diverse model providers through standardized interfaces that abstract away provider-specific authentication, request formatting, and response parsing. The platform currently integrates Azure AI Foundry for EU-hosted models including GPT-4o, GPT-4o-mini, and Mistral variants with guaranteed Swedish or German datacenter processing. We integrate GreenPT for EU-based hosting in datacenters powered by renewable energy. We connect directly to Anthropic for Claude model family access and to OpenRouter for models unavailable through Azure. Each provider relationship involves distinct authentication mechanisms, API patterns, and operational characteristics, but the gateway presents unified interfaces to the frontend.

Adding new providers requires gateway configuration and updating the frontend configuration file, but no changes to user-facing code. When new models emerge that our governance process determines warrant institutional access, we configure gateway connections and update model availability through configuration rather than code modifications. When providers deprecate models, we adjust gateway routing and configuration without touching frontend implementation. This separation of concerns enables the system to evolve with the rapidly changing AI landscape without requiring comprehensive redevelopment of the user interface.

API access is managed through scoped gateway keys rather than sharing raw
vendor credentials. The platform authenticates to Portkey with a master
key capped by budget limits, and we manually issue per-group or per-project keys (through an admin interface)
with tailored permissions, routing rules, and spending ceilings so
course cohorts, research teams, or individual investigators get the
models they need without jeopardising institutional budgets.

Scoped keys prevent the risks of shared credentials: uncontrolled costs,
unapproved routing, and invisible attribution. The gateway logs attach every
call to the right user or group, keeping financial and compliance exposure
manageable. Budget enforcement operates at multiple levels—per-provider caps,
per-model limits, and user or project allocations—preventing runaway spend
while accommodating intensive work, with ceilings adjusted as real usage data
comes in.

\begin{figure}[htbp]
\centering
\includegraphics[width=0.85\textwidth,keepaspectratio]{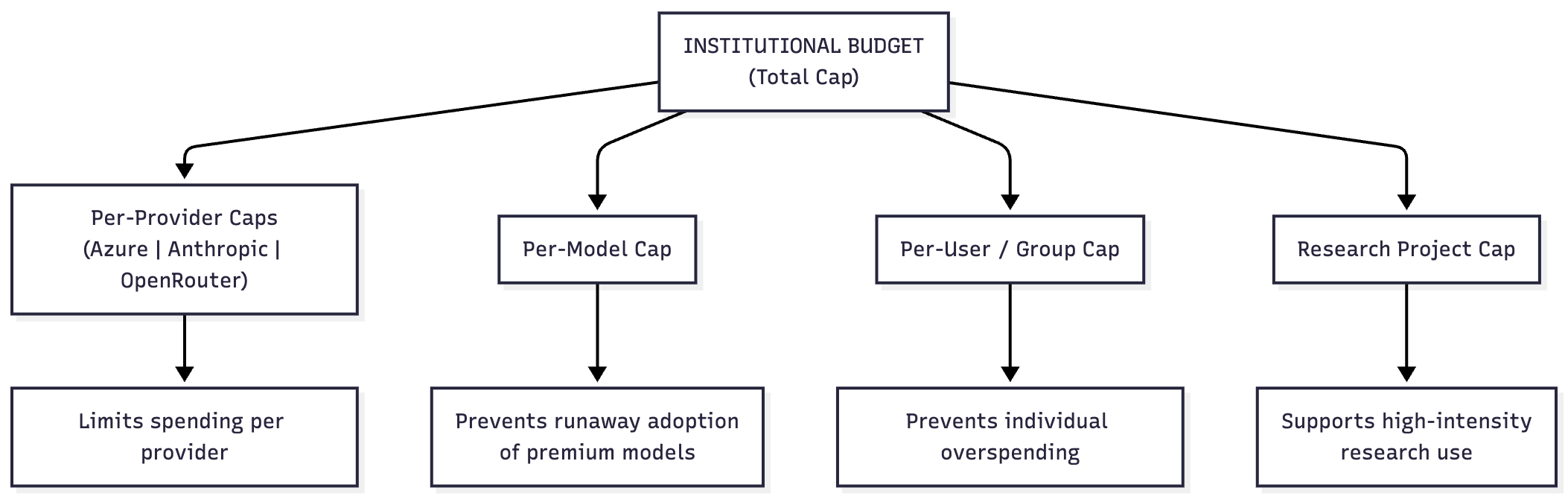}
\caption{Multi-tier budget control and usage monitoring dashboard}
\label{fig:budget-dashboard}
\end{figure}

Dashboards show spending, activity, model utilisation, and cost
projections in real time, broken down by user, model, and period. That
visibility lets us spot valuable or problematic spikes, top up budgets
before they run dry, and ground access or procurement decisions in
observed behaviour when considering new models or cost reductions.

\begin{figure}[htbp]
\centering
\includegraphics[width=0.7\textwidth,keepaspectratio]{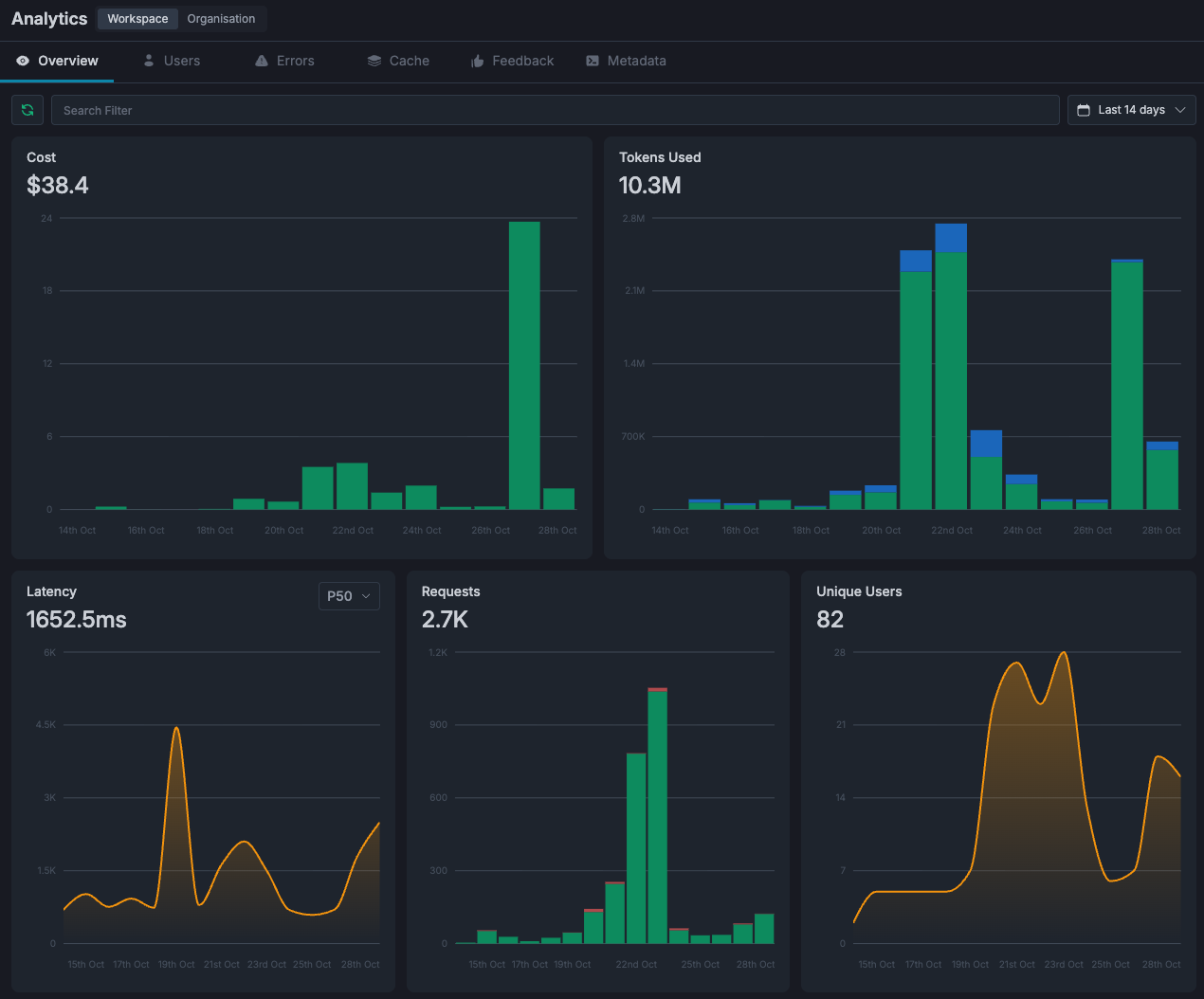}
\caption{Portkey AI gateway administrative interface}
\label{fig:portkey-dashboard}
\end{figure}

\begin{figure}[htbp]
\centering
\includegraphics[width=0.7\textwidth,keepaspectratio]{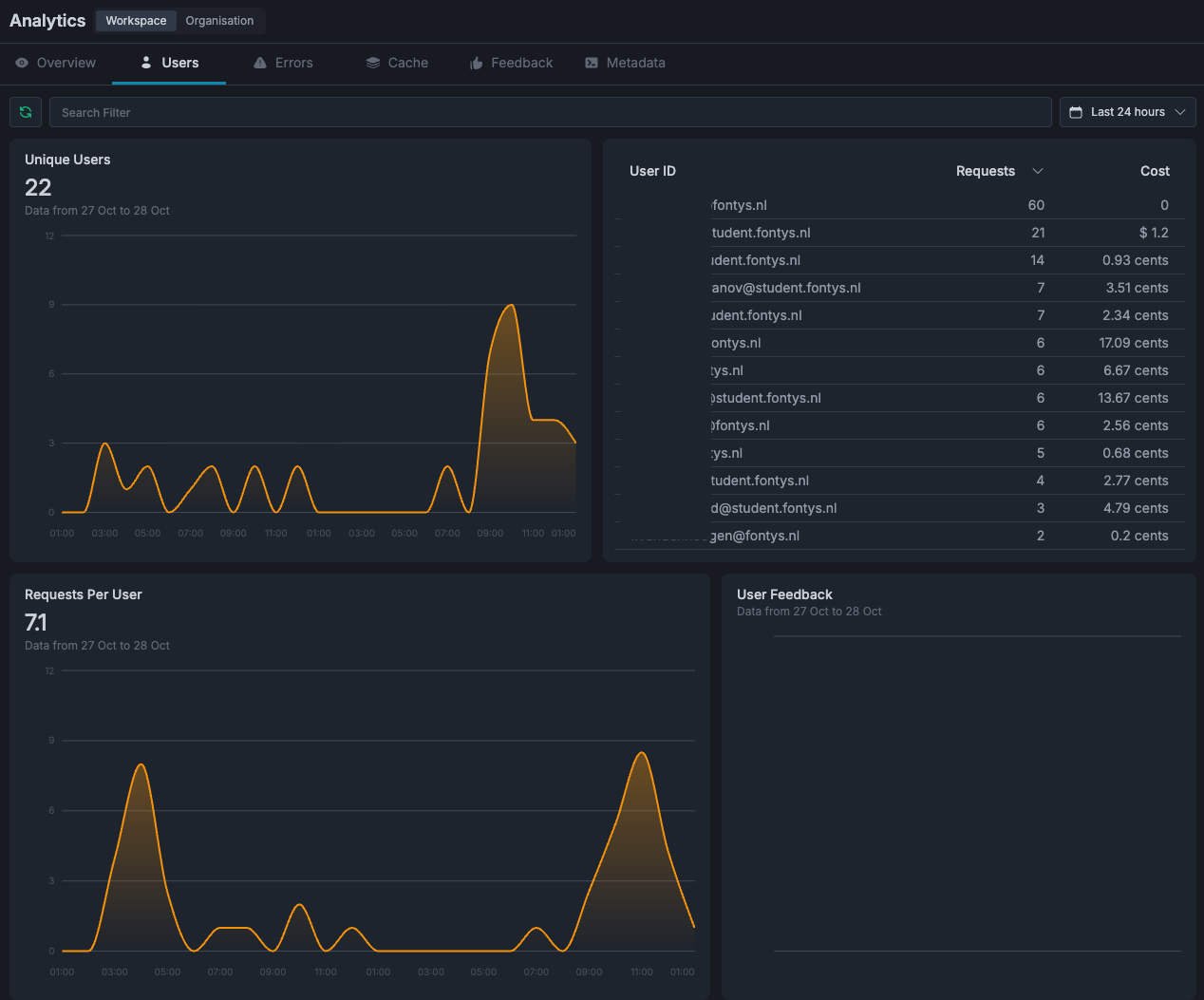}
\caption{Cost breakdown dashboard by user and model}
\label{fig:cost-dashboard}
\end{figure}

Geographic hosting preferences prioritise EU infrastructure. GPT-4o routes through Azure's Swedish region, Mistral models through Azure Germany, and GreenPT serves open-source models from Scaleway's renewable-powered datacenters in France. These defaults keep most traffic within European jurisdiction. When only US hosting exists—as with Anthropic's Claude family—the interface flags the location, requires first-use acknowledgement, and records that acceptance so privacy trade-offs remain explicit rather than accidental.

Access restrictions translate governance decisions into access control. Models with higher privacy risk (non-EU hosting), premium cost profiles, opaque training data, or advanced capability requirements stay hidden until authorised. Claude Sonnet 4, for instance, combines US-only hosting, premium pricing, and limited transparency, so it demands explicit approval.

Users learn about restricted models through the Canvas course accompanying the platform. Requests arrive via support tickets that state the intended use and acknowledge privacy implications. After review, we add approved users to sanctioned Active Directory groups; the gateway enforces those memberships so even technically savvy users cannot bypass the restriction.

In practice, the gateway turns policy into behaviour. EU-first routing, spend caps, and access rules apply automatically, not as guidance users might forget. Governance committees decide the thresholds; the infrastructure ensures compliance without relying on individual discipline.

\subsubsection{External Provider Layer: Provider Selection and Model
Catalog}\label{external-provider-layer-provider-selection-and-model-catalog}

The gateway connects to multiple external providers we vet for hosting location, data handling, and compliance commitments. Core principle: we must always know—and show—where each model runs so GDPR, EU AI Act, and institutional transparency requirements remain satisfied.

Azure AI Foundry anchors the catalog with guaranteed Swedish or German processing for GPT-4o, GPT-4o-mini, Mistral Large, Mistral Medium 3, and an extensive HuggingFace portfolio. These EU-hosted defaults provide capability without sacrificing compliance, so they form baseline institutional access.

GreenPT covers sovereignty-first use cases via Scaleway's renewable-powered French datacenters, including heat recovery for district heating. We expose their self-hosted open-source models (Mistral Mini, GPT-OSS 120B) through OpenWebUI and as direct API endpoints for students who need programmatic access.

For workloads involving sensitive educational data or industry partnerships, we route through the gateway to models running on Fontys-operated hardware in the Netherlands. The gateway logs metadata (timestamps, token counts) for governance, while inference happens entirely on institutional servers to keep conversational content under direct control.

Commercial partners such as Anthropic and OpenAI supply frontier capabilities that still matter for curriculum relevance and research. Because these offerings entail US processing, they automatically invoke the heightened authorisation flow outlined earlier so every cross-border trade-off is documented and acknowledged before use.

Taken together, the catalog balances privacy preferences, capability breadth, and explicit transparency. US hosting never disqualifies a model by default; it simply becomes a conscious choice backed by informed consent and governance oversight.

\begin{figure}[htbp]
\centering
\includegraphics[width=0.8\textwidth,keepaspectratio]{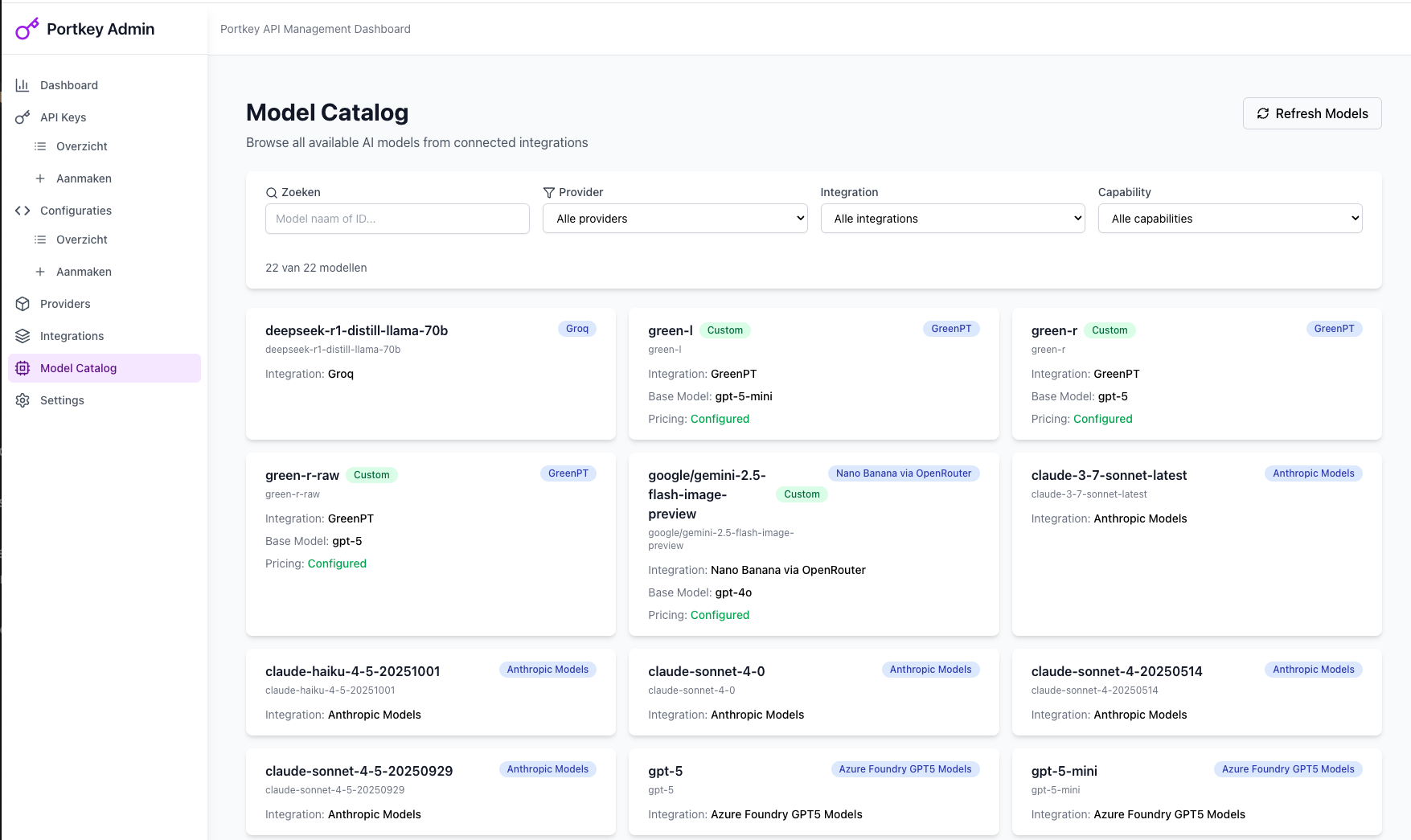}
\caption{Model portfolio showing available AI models including Llama variants (8B, 70B, 405B), Mistral offerings (Small, Medium, Large), and other state-of-the-art open models with performance rivaling proprietary alternatives}
\label{fig:models}
\end{figure}

\subsection{\texorpdfstring{\textbf{Model Cards as Governance
Interface}}{Model Cards as Governance Interface}}\label{model-cards-as-governance-interface}

\subsubsection{From Technical Documentation to Institutional
Decision-Making}\label{from-technical-documentation-to-institutional-decision-making}

Model cards began as transparency artefacts for ML practitioners. When
Margaret Mitchell and colleagues introduced the concept \cite{mitchell2019model}, they proposed a
standardised template so engineers could document provenance, fitness for
purpose, and failure modes. Platforms such as HuggingFace normalised the
format by publishing architectures, training procedures, evaluation
metrics, and known limitations aimed squarely at implementers.

As generative AI shifted into mainstream products, providers repackaged
these documents as compliance-focused system cards: glossier summaries
that mix technical specifications with safety testing, content policies,
and risk mitigations. They satisfy regulators and support brand
protection, but prioritise legal defensibility over developer guidance or
institutional decision support.

Neither format adequately serves educational institutions with
governance responsibilities extending beyond individual technical
implementation or consumer safety. We needed something different:
model cards as active governance instruments serving three
simultaneous functions:

\begin{enumerate}
\def\labelenumi{\arabic{enumi}.}
\tightlist
\item
  \textbf{Systematic risk evaluation} supporting institutional access
  decisions
\item
  \textbf{Transparent documentation} enabling informed user choices
  about which models to use for specific applications
\item
  \textbf{Shared reference framework} facilitating productive
  discussions between users requesting access and governance teams
  evaluating those requests
\end{enumerate}

This reimagining makes transparency serve compliance and education at the
same time. By surfacing each model's characteristics, limitations, privacy
implications, and recommended uses, we turn selection moments into guided
learning experiences. Students confront cost-performance trade-offs,
privacy considerations, and capability boundaries rather than treating AI
as a black box. Information exposure alone does not guarantee better
decisions, but repeated interaction with these cues builds transferable
evaluation habits—an outcome we still plan to validate empirically.

\subsubsection{Addressing the Institutional Documentation
Gap}\label{addressing-the-institutional-documentation-gap}

Commercial system cards nominally meet transparency obligations, yet the
details remain scattered across privacy policies, terms of service, and
technical appendices written for liability protection rather than user
comprehension. Students and faculty rarely have the time—or the expertise—to
parse those sources for each tool. Classic HuggingFace-style cards provide
technical depth, but assume developer control over deployment, which
institutional users simply do not have. The burden of privacy evaluation
therefore falls on individuals instead of sitting within coordinated
governance.

Our framework consolidates the scattered material into standardised
documentation the institution reviews once and then exposes consistently.
Adding a model triggers a structured assessment of:

\begin{itemize}
\tightlist
\item
  Provider documentation and system cards
\item
  Third-party research and benchmarks
\item
  Compliance certifications and audit reports
\item
  Technical specifications and testing results
\end{itemize}

This evaluation produces a model card documenting everything relevant to
decision-making: where data processes geographically, what gets logged
and for how long, what limitations affect output reliability, what costs
accumulate through usage, what alternative models provide similar
capabilities with different trade-off profiles. Users accessing this
consolidated information have the foundation to make better-informed
decisions than they could through scattered individual research.

\subsubsection{From Ethical Requirements to Verifiable
Documentation}\label{from-ethical-requirements-to-verifiable-documentation}

We developed our model card structure by systematically mapping to the
ethical requirements taxonomy proposed by \cite{puhlfuress2025}.
Through thematic analysis of twenty-six AI ethics guidelines, three
documentation frameworks, and ten actual model cards, they identified
forty-three distinct ethical requirements essential for responsible AI
documentation, organized into four ethical
principles—\textbf{Reliability} (technical robustness and security),
\textbf{Transparency} (traceability and explainability),
\textbf{Empowerment} (user autonomy and informed choice), and
\textbf{Beneficence} (fairness and environmental responsibility)—with
three sub-principles under each.

We adapted these twelve sub-principles to institutional
governance contexts, creating a two-tier documentation structure:

Mandatory Disclosure surfaces the most decision-critical
information in every user interface:

\begin{itemize}
\tightlist
\item
  \textbf{Geography} (where data processes and who has access)
\item
  \textbf{Limitations} (what the model cannot reliably do)
\item
  \textbf{Costs} (financial and resource implications)
\end{itemize}

Full Model Cards provide comprehensive ethical evaluation
across twelve sections that map to Puhlfürß et al.'s
framework while addressing institutional needs:

\begin{enumerate}
\def\labelenumi{\arabic{enumi}.}
\tightlist
\item
  \textbf{Model Name \& Version} - Basic identification
\item
  \textbf{Provider \& Hosting} - Geographic and organizational
  accountability (Reliability: Auditability; Empowerment: Liability)
\item
  \textbf{Technical Specifications} - Capabilities and architecture
  (Transparency: Communication of Capabilities)
\item
  \textbf{Intended Use} - Appropriate applications and contexts
  (Transparency: Communication of Capabilities)
\item
  \textbf{Limitations \& Risks} - Known failure modes and boundaries
  (Transparency: Communication of Capabilities; Reliability: Safety and
  Security)
\item
  \textbf{Training Data} - Provenance and characteristics (Transparency:
  Traceability)
\item
  \textbf{Privacy \& Data Handling} - Data subject rights and processing
  (Reliability: Safety and Security; Empowerment: Consent and Control)
\item
  \textbf{Compliance Status} - Regulatory alignment (Reliability:
  Auditability)
\item
  \textbf{Costs \& Resources} - Financial and computational requirements
  (institutional addition)
\item
  \textbf{Sustainability} - Environmental impact (Beneficence:
  Environmental Beneficence)
\item
  \textbf{Comparable Alternatives} - Decision support across models
  (institutional addition)
\item
  \textbf{Governance Status} - Institutional access policies
  (institutional addition)
\end{enumerate}

This structure addresses the full scope of ethical considerations
identified by Puhlfürß et al. while adding institutional governance
components absent from their research-focused taxonomy—but with a
critical constraint that distinguishes our approach from both commercial
system cards and traditional model documentation.

A full example of an institutional Model Card, illustrating the
structure, disclosure format, and compliance mapping is provided in
appendix A.

\subsubsection{\texorpdfstring{\textbf{Interface Integration: Mandatory
Disclosure and Progressive
Documentation}}{Interface Integration: Mandatory Disclosure and Progressive Documentation}}\label{interface-integration-mandatory-disclosure-and-progressive-documentation}

Model cards serve two complementary functions: institutional governance documentation and user-informed consent. However, users do not see full model cards in the chat interface—presenting comprehensive documentation at every interaction would overwhelm users and fail to meet regulatory requirements for digestible, actionable transparency. EU AI Act and GDPR demand information that users can actually process, not long documents hidden behind an "Accept" button. We therefore implement a two-tier disclosure mechanism: a consent modal (similar to cookie consent notifications) that presents critical risk information at the point of model selection, with full model cards accessible for users who need deeper context. This approach ensures that every model invocation is preceded by explicit acknowledgment of material risks (geography, limitations, costs) while maintaining access to comprehensive documentation when needed, turning governance structure into enforceable informed consent practice.

\subsubsection{Technical Implementation: Model-Specific Consent
Configuration}\label{technical-implementation-model-specific-consent-configuration}

We configure mandatory disclosure requirements at the platform level
through OpenWebUI\textquotesingle s model configuration system. Each
model receives a structured consent configuration specifying which risk
elements require explicit acknowledgment before use. This
configuration-driven approach enables systematic consent management
across our multi-provider platform while maintaining flexibility to
adjust disclosure requirements as models, regulations, or institutional
policies evolve.

The configuration distinguishes between universal mandatory
elements---geographic data processing location, key limitations
affecting output reliability, and cost implications when significantly
above baseline models---and model-specific high-risk elements that
warrant additional explicit acknowledgment. US-hosted models trigger
additional required acknowledgments explicitly stating that prompts
process outside the EU with different data protection standards. Models
with undisclosed training data require acknowledgment that bias sources
and content appropriateness cannot be independently verified,
documenting known evaluation limitations rather than obscuring them.

\subsubsection{User Experience: Consent at Point of
Selection}\label{user-experience-consent-at-point-of-selection}

When users select a model requiring explicit consent from the model
selector interface, the platform immediately presents a modal dialog
blocking access until consent is obtained. This consent interface
appears as a prominent overlay---similar to cookie consent notifications
but with substantively more detailed risk information---preventing users
from proceeding to conversation without engaging with material risk
acknowledgments.

The disclosure format keeps critical information concise while forcing
active engagement with each risk element. For Claude Sonnet 4.0, users see
this modal on first selection:

\begin{center}\rule{0.5\linewidth}{0.5pt}\end{center}

\begin{center}
\includegraphics[width=0.7\textwidth,keepaspectratio]{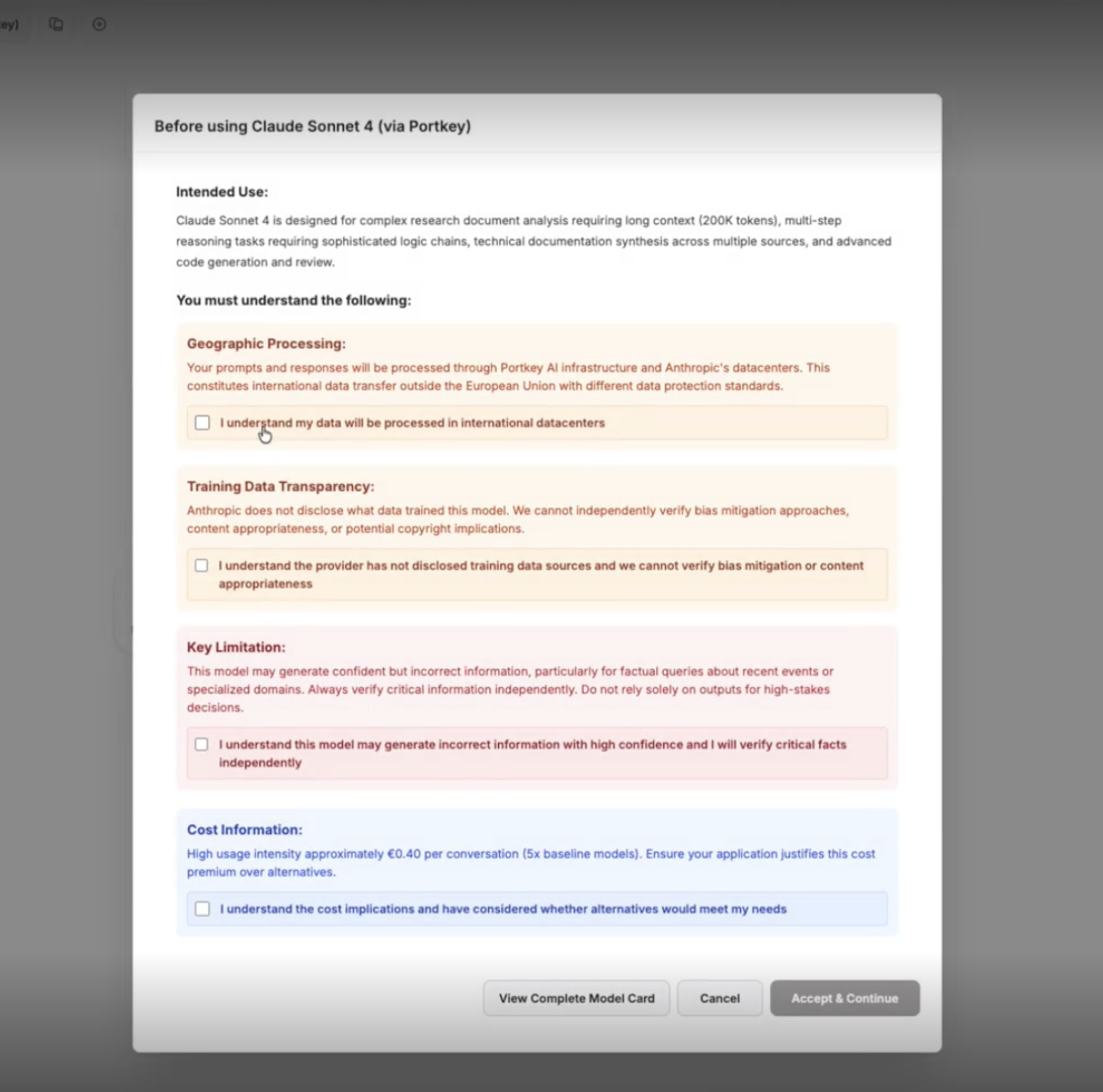}
\end{center}

\begin{center}
\textit{UI implementation of the model card for informed consent}
\end{center}

\begin{center}\rule{0.5\linewidth}{0.5pt}\end{center}

Each checkbox targets a specific risk: geographic processing,
training-data opacity, hallucination likelihood, and cost. Users must
confirm all of them before continuing; no "accept all" shortcut bypasses
the prompts. A prominent link leads to the full model card for anyone who
wants deeper context before granting consent.

Declining the prompt simply returns users to model selection. They either
pick another model or revisit the consent modal when ready to acknowledge
the listed risks.

\subsubsection{Consent Persistence and
Management}\label{consent-persistence-and-management}

OpenWebUI stores consent decisions locally, so users acknowledge each
model once instead of meeting the same dialog every session. A consent
management panel lists granted permissions with dates and lets users
withdraw them; revocation immediately blocks interaction until consent is
renewed. Models without consent remain visible but unusable—opening them
summons the modal again before any prompt can run.

This consent-gating turns GDPR and AI Act transparency into enforced
practice. Every model invocation is preceded by explicit acknowledgment of
geography, limitations, and cost, and the gateway refuses access without
that record.

\subsubsection{Progressive Documentation
Access}\label{progressive-documentation-access}

Full model cards stay one click away for users who need deeper detail when
requesting access, planning sensitive work, or studying system behaviour.
This progressive disclosure keeps everyday use fast while still forcing
everyone to acknowledge core risks, and it supplies auditors with a clear
trail showing that consent is deliberate, recordable, and reversible—not a
box-ticking exercise.

\section{Part III: The Governance Challenge: What Six Months of Operation Revealed}\label{part-iii-the-governance-challenge-what-six-months-of-operation-revealed}

Educational institutions operating their own AI platforms must answer a
governance question that plumbing cannot solve: who decides which models
are offered, on what grounds, for whom, and with what safeguards?

Commercial subscriptions outsource that judgment to vendors. Our gateway
puts budget controls, access rules, and routing preferences in
institutional hands, so the institution must define the decision process
that operationalizes those controls.

We intentionally shipped the technical stack first so the pilot could
surface the real governance work. Some pain points we expected never
materialised; others---model lifecycle, hosting exceptions, access
authorisations, budget steering, compliance evidence, stakeholder
communication---proved central.

This section captures those governance functions, describes how the pilot
handled them, and distils what production-scale operation will require to
serve the wider university.

\subsubsection{\texorpdfstring{\textbf{Model Lifecycle Management:
Addition, Evaluation, and Removal
Authority}}{Model Lifecycle Management: Addition, Evaluation, and Removal Authority}}\label{model-lifecycle-management-addition-evaluation-and-removal-authority}

\subsubsection{\texorpdfstring{\textbf{Operational Questions
Encountered}}{Operational Questions Encountered}}\label{operational-questions-encountered}

The pilot constantly reassessed its catalogue. We opened with Mistral
Large, Claude Sonnet, and Llama 3.3, then new releases (Claude 4.x,
GPT-4o/GPT-5 on Azure, Groq integrations) and niche requests for
embedding, image, or speech models triggered fresh evaluations.

Triggers came from both sides: internal monitoring spotted provider
updates, while users flagged gaps---for example, a research group needing
bge-m3, which we covered by deploying Cohere Embed. Provisioning usually
finished within hours, keeping researchers on the latest models.

We balanced responsiveness with institutional guardrails: favour EU or
open-source hosting when possible, yet provide state-of-the-art options
when pedagogy or research demanded them. Decisions stayed collaborative
but informal; no documented criteria or approval chain existed.

\subsubsection{\texorpdfstring{\textbf{Current Pilot
Approach}}{Current Pilot Approach}}\label{current-pilot-approach}

The technical team executed lifecycle changes, integrating new models
within hours when Portkey support existed. We weighed provider
reliability, cost, and educational or research value, but without
standardised metrics.

Additions and removals were pragmatic:

\begin{itemize}
\tightlist
\item
  \textbf{Added:} \emph{Claude Sonnet 4.5} was deployed one day after
  release to test rapid integration capability and demonstrate platform
  agility.
\item
  \textbf{Reconfigured:} \emph{DeepSeek v1} initially ran via Azure AI
  Foundry but caused severe latency due to hardware limitations;
  migrating it to Groq resolved performance issues and
  underscored the role of infrastructure matching in model selection.
\item
  \textbf{Removed:} \emph{Mistral Small} was discontinued after proving
  less cost-effective than Mistral Medium and Large.
\end{itemize}

Communication about available models matured over time---from implicit
updates to explicit postings in the Teams channel and
Canvas course, allowing users to understand which models were
available for access requests.

\subsubsection{\texorpdfstring{\textbf{What We
Learned}}{What We Learned}}\label{what-we-learned}

Informal, engineer-led choices worked for a small cohort but would crack
under scale. Team members weighted criteria differently, no authority
spoke for institutional priorities, and rapid adoption created hidden
maintenance, support, compliance, and budgeting work. Users meanwhile
assumed availability decisions reflected educational and ethical goals,
not operational convenience; without explicit criteria, transparency
cannot grow with demand.

\subsubsection{\texorpdfstring{\textbf{Production
Requirements}}{Production Requirements}}\label{production-requirements}

Production governance should establish:

\begin{itemize}
\tightlist
\item
  Documented evaluation criteria defining capability
  thresholds, compliance standards, cost ceilings, and pedagogical
  alignment.
\item
  Designated decision authority, such as an AI Governance
  Committee or AI Officer, empowered to approve additions and removals.
\item
  Scheduled reviews to reassess existing models against
  evolving standards and provider changes.
\end{itemize}

A core instrument for managing these functions will be the Model
Card. When a new model is proposed for inclusion, a model card will be
created to document its purpose, capabilities, compliance
characteristics, cost profile, and educational relevance. This card will
serve two governance roles:

\begin{enumerate}
\def\labelenumi{\arabic{enumi}.}
\tightlist
\item
  Decision-making reference --- providing a structured basis
  for evaluating whether the model aligns with institutional criteria
  before integration.
\item
  User transparency --- informing staff, students, and
  researchers about each model's characteristics, limitations, and
  compliance context.
\end{enumerate}

Creating and approving model cards constitutes a governance
responsibility rather than a purely technical task. Someone must own
the drafting of each card, and an authorized body must decide---based on
the documented information---whether that model becomes part of the
institutional portfolio.

Lifecycle management must therefore evolve from opportunistic evaluation
to a systematic process ensuring that model availability decisions are
transparent, traceable, and aligned with institutional objectives.

\subsubsection{\texorpdfstring{\textbf{Geographic Hosting Decisions:
Balancing Data Sovereignty and
Capability}}{Geographic Hosting Decisions: Balancing Data Sovereignty and Capability}}\label{geographic-hosting-decisions-balancing-data-sovereignty-and-capability}

\subsubsection{\texorpdfstring{\textbf{Operational Questions
Encountered}}{Operational Questions Encountered}}\label{operational-questions-encountered-1}

The pilot enforced an EU-first routing philosophy yet kept meeting the
same tension: some projects needed US-only models—often Anthropic’s
Claude—while others deliberately chose EU-hosted options such as Mistral
for privacy. The deciding question quickly became not \emph{where} data
should live, but \emph{when} an exception is justified.

\subsubsection{\texorpdfstring{\textbf{Current Pilot
Approach}}{Current Pilot Approach}}\label{current-pilot-approach-1}

Routing decisions were handled by the technical team. Each case involved
informal evaluation of:

\begin{itemize}
\tightlist
\item
  whether EU-based models could meet the use case,
\item
  the sensitivity of the data processed, and
\item
  the requesting user's ability to understand and manage international
  data transfer implications.
\end{itemize}

This pragmatic approach worked at pilot scale. Users demonstrating clear
research or educational need and awareness of privacy trade-offs
typically received access approval to US-hosted models.

\subsubsection{\texorpdfstring{\textbf{What We
Learned}}{What We Learned}}\label{what-we-learned-1}

Experience confirmed that sovereignty cannot be absolutist: EU-only
hosting would block valid teaching and research, while laissez-faire
exceptions would burn GDPR accountability. The differentiator is
documented, informed consent. When users opt into US-hosted models after
reviewing cross-border implications, institutional values are preserved;
silent routing erodes them.

Model cards are therefore governance instruments as much as reference
material. They record hosting location, compliance posture, and privacy
risks so consent is grounded in facts rather than assumptions.

Our consultation process struck the right balance in principle but added
delay and inconsistency. Production cannot rely on ad-hoc conversations with
stakeholders.

\subsubsection{\texorpdfstring{\textbf{Production
Requirements}}{Production Requirements}}\label{production-requirements-1}

Scaling geographic hosting governance requires a documented framework
specifying:

\begin{itemize}
\tightlist
\item
  Criteria for verifying adequate EU-based alternatives,
\item
  Data sensitivity tiers governing permissible hosting
  locations,
\item
  User awareness thresholds defining who may authorize
  international data transfers, and
\item
  Approval workflows with predictable timelines.
\end{itemize}

Authority for final decisions should rest with a clearly designated
body---either the AI Governance Committee or a privacy-AI
liaison---rather than informal consensus.

Model cards should be formally integrated into this process as both the
information vehicle and the record of decision. Each
model's hosting profile, compliance evaluation, and associated consent
requirements must be clearly described and available to users.

Ultimately, hosting choices reflect institutional value
trade-offs, not technical absolutes. A transparent, documented
framework---anchored in model cards---ensures that these trade-offs are
made deliberately, maintaining both compliance assurance and functional
capability.

\subsubsection{\texorpdfstring{\textbf{Access Authorization: Risk-Based
Model Restrictions and Approval
Workflows}}{Access Authorization: Risk-Based Model Restrictions and Approval Workflows}}\label{access-authorization-risk-based-model-restrictions-and-approval-workflows}

\subsubsection{\texorpdfstring{\textbf{Operational Questions
Encountered}}{Operational Questions Encountered}}\label{operational-questions-encountered-2}

During the pilot, some users needed direct API access rather than the
standard chat interface. The gateway existed for those advanced
educational, research, or development projects that local hardware could
not support, so every API request required formal authorisation even when
the underlying model was available in chat.

Requests followed the \emph{AI Gateway: Direct API Access Guide} on
Canvas and had to show prior local experimentation, a described use case,
faculty or supervisor endorsement, and estimated usage volume.

Submissions ranged from well-scoped thesis projects to curiosity
requests---for example, “please unlock GPT-5” with no accompanying plan.

\subsubsection{\texorpdfstring{\textbf{Current Pilot
Approach}}{Current Pilot Approach}}\label{current-pilot-approach-2}

All access requests were reviewed manually by the technical team. The
same small group handled evaluations and issued project-specific API
keys through Portkey. Decisions were guided by pragmatic criteria:

\begin{itemize}
\tightlist
\item
  Evidence of local AI exploration and a clear justification
  for why gateway access was necessary.
\item
  Use case specificity---educational or research purpose rather
  than casual exploration.
\item
  Faculty or supervisor approval for student-led projects.
\item
  Expected usage pattern and corresponding budget allocation.
\end{itemize}

Requests meeting these standards were typically approved within
one to two business days. Some were denied due to insufficient
justification or missing evidence of local testing. All correspondence
and approvals were retained through email for record-keeping.

\subsubsection{\texorpdfstring{\textbf{What We
Learned}}{What We Learned}}\label{what-we-learned-2}

The approval step did more than cap spend; it forced users to articulate
why they needed high-capability access and discouraged “just unlock the
most expensive model” habits. Yet the informal review style produced
inconsistent thresholds and little audit trail, and recurring questions
showed policy gaps—some students requested access to models already
available, while others glossed over privacy disclosures tied to
US-hosted systems.

\subsubsection{\texorpdfstring{\textbf{Production
Requirements}}{Production Requirements}}\label{production-requirements-2}

Access governance at institutional scale requires standardization and
documentation rather than discretion and memory. Production processes
should establish:

\begin{itemize}
\tightlist
\item
  Structured request forms collecting all relevant information
  consistently---local testing results, project goals, supervisor
  endorsement, and anticipated usage.
\item
  Documented evaluation criteria specifying what constitutes
  adequate justification for various risk and cost levels.
\item
  Defined service-level commitments for response times and
  communication of approval or denial rationale.
\item
  Integration of the Model Card as part of the informed consent
  process---each model's card must disclose data handling location,
  compliance implications, and risk classification, allowing users to
  make informed access decisions.
\end{itemize}

The approval process should remain educational rather than
administrative. When users receive explanations for denials or
redirected approvals, they learn how to select models responsibly and
understand the institutional rationale behind resource governance.

In production, the goal is not to slow users down but to make them
participants in governance---aware of the costs, implications, and
responsibilities of accessing high-capability models within
institutional infrastructure.

\subsubsection{\texorpdfstring{\textbf{Budget Allocation: Cost
Management Beyond Technical
Controls}}{Budget Allocation: Cost Management Beyond Technical Controls}}\label{budget-allocation-cost-management-beyond-technical-controls}

\subsubsection{\texorpdfstring{\textbf{Operational Questions
Encountered}}{Operational Questions Encountered}}\label{operational-questions-encountered-3}

Gateway-level limits stopped runaway spending but raised harder
questions: how much budget sustains a thesis-scale project, how to split
finite funds between universal access and specialised research, and when
to grant top-ups versus nudging users toward cheaper models.

\subsubsection{\texorpdfstring{\textbf{Current Pilot
Approach}}{Current Pilot Approach}}\label{current-pilot-approach-3}

We kept governance simple: every user that requested and was approved for API access received \$10, with a single
justified \$10 top-up available; only one project needed extra shuffling.
The OpenWebUI interface sat under a \$500 monthly institutional cap that
was never breached. Budgets were not class- or faculty-specific, but
usage stayed modest, aided by automated emails warning when users neared
their limits and nudging them to reassess model choices.

\subsubsection{\texorpdfstring{\textbf{What We
Learned}}{What We Learned}}\label{what-we-learned-3}

The pilot showed that technical caps keep spend predictable but cannot
decide how the institution should prioritise resources. More funding for
research projects inevitably means less for broad student access—an
institutional choice, not an engineering one.

Flat allocations kept operations painless yet hinted at the need for
evidence-based differentiation. Most users never hit their limit and only
a single project asked for a top-up, so richer usage insight is still
needed to calibrate allocations that genuinely enable
meaningful educational or research outcomes versus providing mere
convenience.

Transparency for users also proved limited. While individual usage
alerts worked well, students and researchers lacked contextual
benchmarks---no guidance existed on what constituted efficient,
typical, or excessive spending for their type of project. This reduced
the educational potential of cost awareness.

\subsubsection{\texorpdfstring{\textbf{Production
Requirements}}{Production Requirements}}\label{production-requirements-3}

Budget allocation in a production environment must evolve from reactive
adjustments to planned institutional resource management.
Governance should define clear principles for distributing AI
infrastructure funding according to institutional priorities.

Key components include:

\begin{itemize}
\tightlist
\item
  Baseline access ensuring equitable participation for all
  students, independent of personal resources.
\item
  Research allocations scaled to intensity and disciplinary
  need.
\item
  Administrative and operational budgets enabling staff to use
  AI in institutional processes.
\item
  Reserve capacity for experimental programs and emerging use
  cases.
\end{itemize}

Allocation frameworks should be documented and data-informed,
drawing on usage evidence from the pilot and future monitoring. Usage
dashboards and analytics can reveal which models deliver high
educational value per cost, guiding resource rebalancing over time.

Each approved model's Model Card should also include indicative
cost information---typical usage patterns, price-per-token estimates,
and guidance for budget planning. This transparency allows users to
align expectations and plan responsibly before requesting access.

Finally, cost governance should be framed not as restriction, but as
participation. By estimating their needs and reflecting on consumption,
users become co-responsible for sustainable operation of the
institutional AI infrastructure---an essential skill in both digital
literacy and ethical resource use.

\subsubsection{\texorpdfstring{\textbf{Compliance Monitoring: Systematic
Provider Policy Review and Incident
Response}}{Compliance Monitoring: Systematic Provider Policy Review and Incident Response}}\label{compliance-monitoring-systematic-provider-policy-review-and-incident-response}

\subsubsection{\texorpdfstring{\textbf{Operational Questions
Encountered}}{Operational Questions Encountered}}\label{operational-questions-encountered-4}

We treated GDPR and AI Act alignment as foundational. For every new provider
we added, we reviewed provider terms, hosting, and data handling, though certainty was
always limited by the opacity of commercial documentation.

The harder problem was preserving \emph{confidence} over time. Policy
changes from Anthropic, Azure, or OpenAI could have altered retention,
hosting, or acceptable-use terms; none did during the pilot, but the
exercise underlined how fragile manual monitoring is.

\subsubsection{\texorpdfstring{\textbf{Current Pilot
Approach}}{Current Pilot Approach}}\label{current-pilot-approach-4}

Monitoring stayed informal and reactive. A suspected policy change or
user concern triggered a manual review; there was no schedule, automation, or clear owner.

Despite that, compliance awareness guided daily decisions---particularly
model selection and hosting choices. The Model Card provided a
consistent documentation structure, recording each model's hosting
region, data use statements, and known privacy conditions. This practice
offered a partial solution: even without automation, every model
addition triggered at least a minimal structured compliance reflection.

\subsubsection{\texorpdfstring{\textbf{What We
Learned}}{What We Learned}}\label{what-we-learned-4}

The pilot reinforced that GDPR and AI Act alignment is an ongoing
interpretive job. Providers tweak terms without fanfare, and minor text
shifts can carry major implications. Our reactive method sufficed only
because nothing changed.

We also confirmed that compliance needs interdisciplinary ownership:
engineers can trace data flows, but legal and privacy specialists must
judge adequacy. Informal consultation is not sustainable.

Model cards turned abstract compliance into concrete evidence by forcing
every model to declare hosting location, data processing, provider terms,
and documentation quality. Maintaining those cards becomes both the
governance artefact and the audit trail.

\subsubsection{\texorpdfstring{\textbf{Production
Requirements}}{Production Requirements}}\label{production-requirements-4}

Scaling beyond the pilot requires moving from reactive monitoring to
structured oversight. Someone must own responsibility for tracking
provider policy changes—not as crisis response but as continuous
practice. The specific mechanisms will vary by institution, but the
function itself cannot remain informal or ad-hoc.

The Model Card provides the operational foundation for this oversight.
Every update to a provider's terms or capabilities should trigger
revalidation and reissuance of the affected card, creating a continuous
audit trail that links technical configuration to legal accountability.
This discipline aligns practical governance with the evolving
requirements of the EU AI Act.

Incident response also requires structured process. Security breaches,
provider misuse, or user-reported harms need clear notification paths
and escalation procedures connecting technical teams, privacy officers,
and leadership. Without designated ownership and explicit workflows,
compliance monitoring remains vulnerable to gaps and inconsistencies.

\subsubsection{\texorpdfstring{\textbf{Stakeholder Communication:
Transparency About AI Governance and Policy
Evolution}}{Stakeholder Communication: Transparency About AI Governance and Policy Evolution}}\label{stakeholder-communication-transparency-about-ai-governance-and-policy-evolution}

\subsubsection{\texorpdfstring{\textbf{Operational Questions
Encountered}}{Operational Questions Encountered}}\label{operational-questions-encountered-5}

Piloting made it clear that transparency about AI governance matters as
much as functionality. Stakeholders needed tailored context:

\begin{itemize}
\tightlist
\item
  \textbf{Students} needed clarity about which models were available,
  how to request access, how usage affected budgets, and how to report
  issues.
\item
  \textbf{Faculty} required information about pedagogical integration,
  course-level provisioning, and the implications of model selection for
  student data.
\item
  \textbf{Leadership} was indirectly connected through the project's
  inclusion in one of the institutional strategic plan's action lines,
  linking governance development to strategic objectives around digital
  innovation and responsible AI adoption.
\item
  \textbf{Privacy officers} required documentation of data flows,
  consent mechanisms, and incident response procedures.
\item
  \textbf{Researchers and developers} needed technical integration
  details, API documentation, and budget allocation guidance.
\end{itemize}

Each of these groups demanded communication that matched its
responsibilities and perspective---operational, educational, strategic,
or regulatory.

\subsubsection{\texorpdfstring{\textbf{Current Pilot
Approach}}{Current Pilot Approach}}\label{current-pilot-approach-5}

Communication grew organically. The Canvas module hosted documentation,
procedures, and model cards; a Teams channel gave early adopters updates
and a feedback loop; newsletters announced the platform’s existence.
Beyond that, access approvals, ad-hoc emails, and informal faculty
discussions filled gaps, while technical documentation lived on an
internal wiki disconnected from governance messaging.

What we lacked was a coordinated plan. Stakeholders knew the service
existed, but the rationale behind availability decisions, budget
structure, and compliance approach was far less visible.

\subsubsection{\texorpdfstring{\textbf{What We
Learned}}{What We Learned}}\label{what-we-learned-5}

The pilot confirmed that communication must address more than
functionality. Users and leadership alike want to understand
why decisions are made, how trade-offs are balanced,
and what safeguards exist. Without visible governance context,
decisions can appear arbitrary or purely technical.

Different audiences also need different formats: students prefer concise
instructions, faculty want pedagogical framing, leadership expects
strategic synthesis, and the privacy office needs structured evidence. No
single news update covers all of them.

Although no major misunderstandings occurred, the pilot highlighted that
trust depends on predictability and transparency---qualities
that ad-hoc communication cannot sustain.

Students already influence governance indirectly: their access requests for
project-specific models shape what the platform offers, creating emergent
co-creation through use patterns, but research suggests formalizing this
further \cite{chan2023} by involving students directly in developing
acceptable use policies, evaluating which experimental models to support,
and defining what transparency requirements make sense in practice. Making
explicit what is currently implicit would strengthen both policy legitimacy
and student understanding of governance trade-offs.

\subsubsection{\texorpdfstring{\textbf{Production
Requirements}}{Production Requirements}}\label{production-requirements-5}

Scaling governance needs role-specific channels under clear ownership.
Key elements include:

\begin{itemize}
\tightlist
\item
  \textbf{Student-facing documentation} that provides concise guidance
  on model use, access requests, and responsible practices.
\item
  \textbf{Faculty engagement} through workshops and course-integration
  support, aligning AI infrastructure with learning objectives.
\item
  \textbf{Leadership briefings} summarizing governance decisions, cost
  trends, compliance posture, and contribution to the institutional
  strategic plan.
\item
  \textbf{Privacy coordination} via regular compliance updates and
  incident reporting alignment.
\item
  \textbf{Research and technical documentation} integrated into an
  accessible, version-controlled repository linked from the main
  platform.
\end{itemize}

Policy development should include stakeholder consultation
before implementation. Announcing decisions after they are made creates
resistance; involving users early fosters understanding and shared
ownership of governance choices.

Regular updates---preferably on a quarterly cadence---should
provide stakeholders with concise summaries of platform status, usage
trends, governance developments, and upcoming changes. These updates
should be institutional artifacts, not just announcements: they document
active governance and make oversight visible.

Finally, communication should not be one-directional. \textbf{Feedback
mechanisms} through Teams channels, structured surveys, or course-level
reflection sessions should allow users to raise issues and suggest
improvements. The institution should close the loop by reporting back on
what input was received and how it shaped subsequent decisions.

Transparent communication transforms AI governance from an internal
administrative process into a visible institutional practice. It
demonstrates that the infrastructure is managed not just \emph{for}
users, but \emph{with} them---anchoring trust, accountability, and
educational alignment as core elements of sustainable AI operation.

\subsubsection{\texorpdfstring{\textbf{Institutional Capacity for
Production-Ready
Governance}}{Institutional Capacity for Production-Ready Governance}}\label{institutional-capacity-for-production-ready-governance}

The governance functions described above---model lifecycle management,
geographic hosting decisions, access authorization, budget allocation,
compliance monitoring, and stakeholder communication---share common
characteristics that extend far beyond technical administration.
Together they define the institutional capability required to operate AI
infrastructure as a sustainable educational and research service.

\begin{enumerate}
  \item \textbf{Cross-functional coordination} is essential. Effective
governance spans IT infrastructure, educational strategy, privacy and
legal compliance, research facilitation, and institutional leadership.
No single existing role naturally combines these perspectives;
structured collaboration is required.

  \item \textbf{Continuous attention} replaces one-time setup. AI governance is
a living process shaped by changing provider policies, new model
capabilities, evolving regulation, and shifting institutional
priorities. Research on AI policy in higher education confirms this
operational reality: governance frameworks must emphasize ``continuous
improvement and adaptation, enabling universities to refine their AI
integration strategies in response to new insights and changing needs''
\cite{chan2023}. Processes that are not continuously reviewed and adapted
will rapidly become obsolete.

  \item \textbf{Institutional authority} must accompany technical capability.
Decisions about which models are available, where data is processed, and
how budgets are distributed cannot rely solely on operational
discretion. They must carry organizational legitimacy, reflect
educational values, and align with the institutional strategic plan.

  \item \textbf{Specialized expertise} underpins credible governance.
Sustainable oversight requires a blend of technical understanding of AI
systems, familiarity with regulatory obligations such as the GDPR and
the EU AI Act, and insight into how AI integrates into teaching and
research practice. This combination is rare and cannot be assumed to
exist implicitly within existing teams.
\end{enumerate}

Based on six months of operational experience, these requirements point
to the need for designated institutional capacity.

The Model Card offers a practical foundation for this
capacity---an instrument connecting technical, educational, and
compliance perspectives into a single traceable artifact. Each model
card documents capabilities, hosting, privacy characteristics, cost
implications, and educational relevance, providing a unified reference
for governance decisions and user transparency.

Institutional AI infrastructure becomes sustainable not through
technical maturity alone, but through governance that is coordinated,
accountable, and visible. Production readiness means more than reliable
systems---it means a governance framework capable of learning, adapting,
and demonstrating that institutional control of AI is both operationally
effective and socially responsible.

\subsubsection{\texorpdfstring{\textbf{The AI Officer Role: A New
Institutional
Function}}{The AI Officer Role: A New Institutional Function}}\label{the-ai-officer-role-a-new-institutional-function}

The governance functions described above require an institutional role
that bridges technical infrastructure, educational purpose, and
regulatory compliance. We therefore propose establishing an AI Officer, 
reporting to institutional leadership, with explicit responsibility for 
systematic AI governance, model portfolio oversight, and institutional 
coordination.

This role is not an extension of existing IT or privacy functions. It
represents a new institutional capacity---the first point where
technical understanding, governance authority, and educational
responsibility converge. The AI Officer ensures that institutional AI
remains technically robust, legally compliant,
financially sustainable, and pedagogically aligned.

\subsubsection{\texorpdfstring{\textbf{Why This Must Be a Dedicated
Role}}{Why This Must Be a Dedicated Role}}\label{why-this-must-be-a-dedicated-role}

AI governance cannot function as an informal coordination task. Our
pilot operation demonstrated that decisions about model inclusion,
access authorization, compliance review, and budget allocation require
continuous attention, not occasional involvement.

This need reflects a fundamental shift in how AI relates to institutional strategy. Unlike conventional technologies, AI is not merely a support function to be aligned with strategy---AI is strategy itself. As AI continues to reshape how institutions compete, innovate, and deliver value, responsibility for its deployment and governance cannot remain ad hoc or fragmented \cite{butcher2024}. A new leadership function is required, one that is structurally embedded in institutional leadership and strategically empowered to orchestrate AI across the enterprise.

Technical staff possess the skills to integrate models but lack institutional authority
for value-based governance. Administrators and privacy officers possess
governance expertise but cannot assess the technical validity of model
claims or provider documentation. This gap is not unique to us:
analysis of executive AI leadership identifies a structural ``executive
gap'' in most organizations---``the absence of a formally recognized
role that can integrate technical knowledge and strategic direction for
AI at the enterprise level'' \cite{butcher2024}. Traditional technology
leaders remain ``reactive, execution-focused, and typically confined to
infrastructure or product domains''; they lack the institutional mandate
for value-based governance.

The AI Officer occupies this intersection, translating institutional
values into operational practice. Without dedicated capacity, governance
becomes fragmented---dependent on whoever happens to notice provider
changes, handle access requests, or respond to compliance concerns. This governance vacuum leads to inconsistencies in deployment, duplication of efforts, and heightened exposure to regulatory and reputational risks \cite{butcher2024}. As
AI systems evolve, this fragility becomes untenable. Governance must
have both institutional backing and operational continuity.

AI systems introduce novel governance challenges that transcend traditional functional boundaries. Unlike conventional technologies, AI models are dynamic, probabilistic, and often non-deterministic. They evolve through learning, exhibit opaque internal logic, and can produce emergent effects not anticipated at design time \cite{butcher2024}. These properties introduce novel forms of ethical, operational, and reputational risk that defy existing accountability structures, demanding a leadership function specifically attuned to AI's unique characteristics.

Production-scale governance also demands positional authority. Decisions
about which models enter the institutional portfolio or how student data
is processed affect multiple stakeholders and must carry organizational
legitimacy. A formal role prevents governance from being interpreted as
personal preference or technical convenience. The AI Officer thus serves
as both operational coordinator and institutional representative for AI
governance.

\subsubsection{\texorpdfstring{\textbf{Required
Competencies}}{Required Competencies}}\label{required-competencies}

The AI Officer blends two scarce skill sets. First, \textbf{technical
literacy}: working knowledge of model architectures, provider ecosystems,
APIs, and performance trade-offs, plus the fluency to read compliance
documentation, spot data-processing implications, and judge when
exceptions are justified. Second, institutional governance
capability: the capacity to craft policies that balance educational
needs, privacy obligations, cost realities, and strategic goals, grounded
in GDPR, the EU AI Act, and educational data ethics, and reinforced by
cross-functional coordination.

Separated, these abilities fail—technical insight without governance
produces unaccountable systems; governance without technical grounding
creates impractical policy. The AI Officer exists to keep them fused.

\subsubsection{\texorpdfstring{\textbf{Responsibility Scope and
Authority
Requirements}}{Responsibility Scope and Authority Requirements}}\label{responsibility-scope-and-authority-requirements}

The AI Officer spans policy design and day-to-day execution to keep AI
governance transparent, consistent, and aligned with educational goals.
Core responsibilities include:

\begin{itemize}
\tightlist
\item  \textbf{Model portfolio governance}: evaluate, approve, and retire
models against institutional criteria and maintain their
Model Cards as the audit trail for every decision.
\item  \textbf{Access authorisation}: run predictable approval processes,
document rationales, and balance capability needs with privacy and cost.
\item  \textbf{Compliance coordination}: work with privacy and legal teams
to schedule provider reviews, maintain processing records, and trigger
incident response when needed.
\item  \textbf{Educational and research integration}: guide faculty and
researchers on appropriate use, model behaviour, and privacy impacts.
\item  \textbf{Budget planning and reporting}: align allocations with
institutional priorities, analyse usage data, and advise on financial
implications of new models or providers.
\end{itemize}

Reporting to institutional leadership gives the
AI Officer legitimacy to approve or decline integrations, enforce
standards, and recommend resource allocations.

\subsubsection{\texorpdfstring{\textbf{Organizational Implementation:
Individual Role or Shared
Responsibility}}{Organizational Implementation: Individual Role or Shared Responsibility}}\label{organizational-implementation-individual-role-or-shared-responsibility}

This function can be implemented as a single role or a small,
coordinated team. Whichever structure wins, the test is capacity: the
institution needs sustained expertise that can cover portfolio, access,
compliance, education, and budgeting.

A single officer offers clear accountability and consistent application
of criteria. A distributed governance team can share specialist
responsibilities while acting as one institutional voice. In either
case, the function must be formally designated, recognised, and
empowered to approve or reject integrations, maintain model cards,
coordinate compliance reviews, and advise leadership on budget and
strategy. Anything less leaves governance to informal initiative rather
than institutional commitment.

\section{Conclusion: From Pilot to Principle}\label{conclusion-from-pilot-to-principle}

The six-month pilot began as an engineering test and ended as proof that
architecture is governance. Every decision about routing, access, and
transparency encoded values; building our own gateway meant we could no
longer outsource responsibility.

We showed that institutional AI sovereignty is viable when governance
receives the same design discipline as code: equitable access, privacy
standards, and educational autonomy can coexist. The deliverable is not a
product but a configurable, value-driven framework.

Scaling such platforms depends on institutional maturity. Governance must move from
good intentions to structured capacity: someone responsible for tracking provider changes,
systematic oversight instead of crisis response, processes that adapt as AI evolves.
Trust comes from consistent accountability, not comprehensive procedures.

This requires dedicated leadership for a fundamental reason: AI is no longer a support
function but strategy itself. Systems that learn, evolve, and produce unexpected effects
demand governance beyond traditional IT or privacy functions. The AI Officer bridges
infrastructure, regulation, and pedagogy—treating AI as the strategic foundation it has
become, not merely an operational tool to manage.

The lesson extends beyond any single institution. As AI becomes central to how
institutions operate, public education and other public institutions cannot depend on
commercial systems built for profit over values. The challenge is universal: AI is
everywhere, yet no one is formally responsible. Dedicated leadership and structured
oversight protect the ability to decide how knowledge is produced, processed, and
safeguarded. Designing our own infrastructure reclaimed agency. The commitment is clear: we should govern AI tools with the same transparency, accountability, and ethics we teach students to practice.

\clearpage
\section{Example Model Card Claude Sonnet 4.5}\label{appendix-a}

\begin{tcolorbox}[colback=gray!10, colframe=gray!80, breakable, title=Model Card -- Claude Sonnet 4.5]

\subsection{1. Model Name and Version}\label{1-model-name-and-version}
\begin{modelcardpanel}
\textbf{Model Name:} Claude Sonnet 4.5

\textbf{Version:} Release Date: September 29, 2024

\textbf{Model Identifier:} \texttt{claude-sonnet-4-5-20250929}

\textbf{Provider:} Anthropic PBC

\emph{This is the current flagship model in Anthropic\textquotesingle s
Claude 4 family, representing the most advanced iteration of Sonnet.
Version tracking matters---Anthropic frequently updates models, and
behavioral characteristics can shift between releases. When referencing
this model in research, technical documentation, or access requests,
always include the full version identifier to ensure clarity.}
\end{modelcardpanel}

\subsection{2. Provider and Hosting
Location}\label{2-provider-and-hosting-location}
\begin{modelcardpanel}
\textbf{Provider Organization:} Anthropic PBC (Public Benefit
Corporation)

\textbf{Primary Hosting:} United States (AWS datacenters in Oregon and
Virginia regions)

\textbf{EU Availability:} None. Anthropic does not currently offer
European datacenter hosting for any Claude models.

\textbf{Data Processing Flow:} All inference requests route directly to
Anthropic\textquotesingle s US infrastructure via AWS. Your prompts and
responses travel internationally---there is no European processing
option regardless of payment tier or special arrangements.

\textbf{Geographic Reality Check:} This means GDPR Article 44
international data transfer considerations apply to every conversation.
Anthropic relies on Standard Contractual Clauses (SCCs) as the legal
mechanism for EU-US data transfers. While SCCs are an accepted
framework, they place compliance demonstration burden partially on the
institution rather than being as straightforward as EU-hosted
alternatives.

\textbf{Why No EU Hosting?} Anthropic has prioritized model capability
and rapid iteration over geographic distribution. They concentrate
infrastructure in US regions where they maintain operational expertise
and can deploy updates quickly. This is a strategic choice---not a
temporary limitation we expect to resolve soon.
\end{modelcardpanel}

\subsection{3. Technical
Specifications}\label{3-technical-specifications}
\begin{modelcardpanel}
\textbf{Architecture Type:} Transformer-based large language model
(constitutional AI training methodology)

\textbf{Parameter Count:} Undisclosed (Anthropic does not publish exact
parameter counts for competitive reasons)

\textbf{Context Window:} 200,000 tokens (\textasciitilde150,000 words or
500+ pages)

\textbf{Maximum Output:} 8,192 tokens per response

\textbf{Supported Modalities:}

\begin{itemize}
\tightlist
\item
  \textbf{Input:} Text, images (PNG, JPEG, GIF, WebP), PDF documents
  (with text extraction)
\item
  \textbf{Output:} Text only (no image generation, though can produce
  code that generates images)
\end{itemize}

\textbf{Inference Speed:} Approximately 60-80 tokens/second for typical
requests (varies with load and complexity). This is middle-tier
speed---faster than Opus, slower than Haiku variants.

\textbf{Notable Technical Characteristics:}

\begin{itemize}
\tightlist
\item
  Extended reasoning mode available through specific API parameters
  (triggers deeper analysis chains)
\item
  Strong function calling support for tool use and structured outputs
\item
  Reliable JSON mode for applications requiring formatted data
\item
  Excels at following complex multi-step instructions with maintained
  context
\end{itemize}

\textbf{What "Sonnet" Means in Anthropic\textquotesingle s Lineup:}
Sonnet represents their balanced model tier---smarter than Haiku
(fast/cheap), more efficient than Opus (maximum capability/expensive).
Think of it as the "Goldilocks" option for most sophisticated
applications that don\textquotesingle t require absolute frontier
performance.
\end{modelcardpanel}

\subsection{4. Intended Use}\label{4-intended-use}
\begin{modelcardpanel}
\textbf{Provider\textquotesingle s Stated Intended Use:} Complex text
analysis, research synthesis, technical writing assistance, code
generation and review, extended document processing, multi-step
reasoning tasks, conversational agents requiring nuanced understanding.

\textbf{Institutional Assessment:}

This model excels in scenarios requiring:

\begin{itemize}
\tightlist
\item
  \textbf{Research document analysis} where you\textquotesingle re
  synthesizing information across multiple 50+ page papers
\item
  \textbf{Complex coding tasks} involving architecture decisions,
  debugging across multiple files, or explaining intricate algorithms
\item
  \textbf{Extended technical writing} where you need to maintain
  consistent voice and logical flow across long-form content
\item
  \textbf{Multi-turn problem-solving} where the conversation builds on
  previous context over dozens of exchanges
\item
  \textbf{Structured thinking} about ambiguous problems requiring
  breaking down complexity systematically
\end{itemize}

\textbf{Educational Level Appropriateness:}

\begin{itemize}
\tightlist
\item
  \textbf{MA-AAI1 Advanced Students:} Ideal for guild projects requiring
  sophisticated reasoning across Architect, Engineer, or Experience
  Designer specializations
\item
  \textbf{Research Applications:} Well-suited for professorship
  investigations requiring nuanced analysis
\item
  \textbf{Undergraduate Students:} Generally appropriate, though
  consider if less expensive EU-hosted alternatives (GPT-4o via Azure,
  Mistral Large) adequately serve the use case before defaulting to this
  premium option
\end{itemize}

\textbf{When NOT to Use This Model:}

\begin{itemize}
\tightlist
\item
  Simple factual lookups (use GPT-4o-mini or Mistral Small
  instead---same accuracy, 1/5th the cost)
\item
  Quick code snippets under 50 lines (cheaper models handle this fine)
\item
  Tasks requiring image generation (this model doesn\textquotesingle t
  generate images)
\item
  Applications with sensitive personal data unless US processing is
  explicitly acceptable
\item
  Casual conversation practice (reserve expensive models for genuine
  complexity)
\end{itemize}
\end{modelcardpanel}

\subsection{5. Limitations and Risks}\label{5-limitations-and-risks}
\begin{modelcardpanel}
\textbf{Hallucination Risk:} Moderate to High for certain query types

Like all large language models, Claude Sonnet 4.5 can generate
authoritative-sounding but incorrect information, particularly for:

\begin{itemize}
\tightlist
\item
  \textbf{Recent events} after training cutoff (January 2025)---it will
  confidently fabricate details about news, research, or events it
  wasn\textquotesingle t trained on
\item
  \textbf{Precise quantitative claims} requiring exact calculation---it
  approximates rather than computes
\item
  \textbf{Niche technical domains} with sparse training
  representation---specialized engineering, emerging technologies,
  obscure academic subfields
\item
  \textbf{Specific citations} it claims to reference---it may invent
  paper titles, authors, or publication details that sound plausible but
  don\textquotesingle t exist
\end{itemize}

\textbf{Critical Verification Requirement:} ALWAYS independently verify
factual claims for high-stakes decisions. Use this model as a
brainstorming partner and analysis tool, not as an authoritative
reference source. Cross-check specific facts, citations, technical
specifications, or numerical claims before relying on them.

\textbf{Training Data Transparency:} None

Anthropic does not disclose:

\begin{itemize}
\tightlist
\item
  What corpora trained this model (web crawl? books? code repositories?
  proprietary datasets?)
\item
  How they curated or filtered training data
\item
  What content they explicitly excluded or included
\item
  Copyright status of training materials
\end{itemize}

\textbf{What This Opacity Means:} We cannot independently verify bias
sources, content appropriateness for educational contexts, or potential
copyright implications in generated outputs. You\textquotesingle re
trusting Anthropic\textquotesingle s internal processes without external
validation. This is an \emph{industry-standard} limitation---most
frontier model providers maintain similar opacity---but that
doesn\textquotesingle t make it less problematic for institutional
assessment.

\textbf{Bias Considerations:}

Third-party evaluations and user reports document several consistent
patterns:

\begin{itemize}
\tightlist
\item
  \textbf{Cultural representation gaps:} Training data skews toward
  Western (particularly American) contexts. Expect weaker performance
  for non-Western cultural references, holidays, social norms,
  historical events, or idioms.
\item
  \textbf{English language dominance:} While multilingual capable,
  quality degrades noticeably in non-English languages. Dutch
  performance is adequate but not native-level---expect occasional
  awkward phrasing or cultural context misses.
\item
  \textbf{Professional stereotyping:} May default to gendered
  assumptions in career contexts (nurses as female, engineers as male,
  etc.) despite training to avoid this
\item
  \textbf{Socioeconomic assumptions:} Training data likely
  over\-represents middle/upper-class perspectives---may not accurately
  reflect lived experiences of students from diverse economic
  back\-grounds
\end{itemize}

\textbf{Recommendation:} Review outputs critically when model generates
content about underrepresented populations, non-Western contexts, or
makes assumptions about user circumstances. The model\textquotesingle s
"default human" tends to be Western, English-speaking, and relatively
privileged.

\textbf{Reasoning Limitations:}

Despite sophisticated capabilities, documented failure modes include:

\begin{itemize}
\tightlist
\item
  \textbf{Novel problem types} outside training distribution---it
  pattern-matches rather than truly reasons, so genuinely unprecedented
  problems may confuse it
\item
  \textbf{Multi-step mathematical proofs} requiring rigorous logical
  chains---it can approximate but may introduce subtle errors in
  extended derivations
\item
  \textbf{Adversarial queries} deliberately designed to exploit model
  weaknesses---it\textquotesingle s not robust against sophisticated
  jailbreaking or manipulation attempts
\item
  \textbf{Self-knowledge:} It doesn\textquotesingle t accurately know
  what it knows---confidence levels don\textquotesingle t reliably
  indicate accuracy
\end{itemize}

\textbf{Context Retention Over Long Conversations:}

Even with 200K token context window, accuracy and coherence degrade
across very long conversations involving multiple distinct topics. The
model maintains focus better than most alternatives, but
you\textquotesingle ll still see:

\begin{itemize}
\tightlist
\item
  Drift in conversational tone or persona after 50+ exchanges
\item
  Forgetting details mentioned early in conversation when context
  approaches token limits
\item
  Confusion when mixing multiple unrelated topics in single session
\end{itemize}

\textbf{Best Practice:} Start fresh conversations for distinct projects
or topics rather than trying to handle multiple subjects in one endless
thread.
\end{modelcardpanel}

\subsection{6. Training Data and
Evaluation}\label{6-training-data-and-evaluation}
\begin{modelcardpanel}
\textbf{Training Data Sources:} Undisclosed

\textbf{Data Curation Methodology:} Undisclosed

\textbf{Evaluation Datasets:} Partially disclosed through
Anthropic\textquotesingle s published research

\textbf{Training Cutoff:} January 2025

\textbf{What Anthropic Has Disclosed:}

Through published research and blog posts, we know:

\begin{itemize}
\tightlist
\item
  Training included "constitutional AI" methodology emphasizing
  helpfulness, harmlessness, and honesty through reinforcement learning
  from AI feedback (RLAIF)
\item
  Data filters aimed to reduce toxic content, personal information, and
  copyrighted material---but specific mechanisms remain proprietary
\item
  Evaluation used standard benchmarks including MMLU, Human\-Eval, GSM8K,
  and internal safety assess\-ments
\end{itemize}

\textbf{Performance Claims from Provider:}

Anthropic reports Claude Sonnet 4.5 achieves:

\begin{itemize}
\tightlist
\item
  89.8\% on MMLU (graduate-level knowledge across 57 subjects)
\item
  94.6\% on HumanEval (Python coding problems)
\item
  96.7\% on GSM8K (grade school math word problems)
\item
  Strong performance on constitutional AI safety metrics
\end{itemize}

\textbf{Institutional Skepticism Warranted:} These are
\emph{provider-reported} figures using datasets and evaluation
methodologies we cannot independently verify. Benchmark performance
doesn\textquotesingle t always predict real-world capability---models
can overfit to common evaluation tasks while struggling with novel
applications. Consider these figures rough capability indicators, not
precise performance guarantees.

\textbf{Subgroup Performance Analysis:} Not published

We have no independent data on:

\begin{itemize}
\tightlist
\item
  Performance differences across demographic groups
\item
  Capability variations across languages beyond English
\item
  Bias metrics for outputs affecting different populations
\item
  Fairness evaluations across protected characteristics
\end{itemize}

This represents a significant documentation gap. Models showing strong
aggregate performance can still exhibit substantial performance
disparities across subgroups, potentially creating equity issues in
educational applications.
\end{modelcardpanel}

\subsection{7. Privacy and Data
Handling}\label{7-privacy-and-data-handling}
\begin{modelcardpanel}
\textbf{What Gets Logged:}

Anthropic retains:

\begin{itemize}
\tightlist
\item
  \textbf{Request metadata:} Timestamps, API keys, token counts, model
  version, error logs
\item
  \textbf{Conversational content:} Your prompts and
  Claude\textquotesingle s responses
\item
  \textbf{Usage patterns:} Frequency, topic clustering for abuse
  detection
\end{itemize}

\textbf{Retention Periods:}

Per Anthropic\textquotesingle s current privacy policy:

\begin{itemize}
\tightlist
\item
  \textbf{Active conversations:} Retained for 30 days for operational
  troubleshooting and abuse prevention
\item
  \textbf{After 30 days:} Conversational content deleted unless you
  explicitly save it or it\textquotesingle s flagged for Trust \& Safety
  review
\item
  \textbf{Metadata:} Retained longer for billing, analytics, and service
  improvement---specific timelines vary by data type
\end{itemize}

\textbf{Training Data Usage:}

Anthropic\textquotesingle s contractual commitment for API usage (which
our gateway uses):

\begin{itemize}
\tightlist
\item
  \textbf{Your conversations do NOT enter training pipelines} for future
  models
\item
  \textbf{Exception:} Content flagged through Trust \& Safety review for
  abuse (violence, illegal content, CSAM) may be retained for safety
  research
\item
  This is a \emph{contractual commitment} rather than technical
  impossibility---we must trust Anthropic honors this agreement
\end{itemize}

\textbf{Geographic Processing:} US-Only

Despite routing through our gateway, your data still processes
exclusively in Anthropic\textquotesingle s US datacenters. Our gateway
adds EU-based routing and budget controls, but cannot change fundamental
data processing geography. The actual model inference happens in Oregon
or Virginia AWS facilities.

\textbf{Audit Mechanisms:}

Limited external verification available:

\begin{itemize}
\tightlist
\item
  Anthropic undergoes SOC 2 Type II audits covering security controls
\item
  No independent GDPR compliance audits publicly available
\item
  No technical verification mechanisms allowing institutions to confirm
  data handling claims
\item
  Relies primarily on contractual terms and Anthropic\textquotesingle s
  reputation
\end{itemize}

\textbf{Institutional Data Processing Agreement:}

We maintain a data processing agreement with Anthropic covering:

\begin{itemize}
\tightlist
\item
  Contractual commitment to 30-day content deletion
\item
  Standard Contractual Clauses for EU-US transfer compliance
\item
  Sub-processor disclosure (AWS as infrastructure provider)
\item
  Limited liability for model outputs
\end{itemize}

\textbf{What We Cannot Verify:}

Despite contractual agreements, we lack technical mechanisms to
independently confirm:

\begin{itemize}
\tightlist
\item
  Whether content actually deletes after 30 days (we trust, cannot
  verify)
\item
  What Anthropic\textquotesingle s internal staff can access and under
  what circumstances
\item
  How abuse detection algorithms work or what triggers content review
\item
  Whether metadata retention serves only stated purposes
\end{itemize}

\textbf{This is standard industry practice} but remains a transparency
gap. We document it clearly rather than pretending certainty we lack.
\end{modelcardpanel}

\subsection{8. Compliance Status}\label{8-compliance-status}
\begin{modelcardpanel}
\textbf{EU AI Act Risk Classification:} General Purpose AI Model (not
high-risk per current interpretation, subject to change as guidance
evolves)

\textbf{GDPR Compliance Verification:}

\begin{itemize}
\tightlist
\item
  \textbf{Legal Mechanism:} Standard Contractual Clauses (SCCs) for
  EU-US data transfer per GDPR Article 46
\item
  \textbf{Verification Method:} Contractual review + legal counsel
  assessment---no independent technical audit
\item
  \textbf{Last Review:} October 2024 (ongoing monitoring)
\item
  \textbf{Adequacy:} Legally permissible under current framework, though
  EU-only hosting would simplify compliance demonstration
\end{itemize}

\textbf{Relevant Certifications:}

\begin{itemize}
\tightlist
\item
  \textbf{SOC 2 Type II:} Achieved (security, availability,
  confidentiality controls verified by third-party auditor)
\item
  \textbf{ISO 27001:} Not claimed by Anthropic
\item
  \textbf{GDPR Certification:} No formal certification (SCCs provide
  legal basis instead)
\end{itemize}

\textbf{Compliance Gaps Requiring Institutional Mitigation:}

\begin{enumerate}
\def\labelenumi{\arabic{enumi}.}
\tightlist
\item
  \textbf{International Transfer Risk:} US processing requires explicit
  user acknowledgment per GDPR Article 49. Our platform implements this
  through mandatory pre-use disclosure.
\item
  \textbf{Training Data Transparency:} GDPR Article 13 requires
  information about data processing purposes and legal basis.
  Anthropic\textquotesingle s opacity about training data sources
  creates documentation gaps we must acknowledge to users.
\item
  \textbf{Automated Decision-Making:} GDPR Article 22 requires
  disclosure when automated processing significantly affects
  individuals. Users must understand that AI outputs
  shouldn\textquotesingle t drive high-stakes decisions without human
  review.
\item
  \textbf{Data Subject Rights:} GDPR grants rights to access,
  rectification, and erasure. Anthropic\textquotesingle s 30-day
  retention supports erasure in principle, but we cannot offer immediate
  deletion upon user request---creates 30-day window where data remains
  accessible to provider.
\end{enumerate}

\textbf{Verification Confidence Level:} Medium

We have:

\begin{itemize}
\tightlist
\item[\(\checkmark\)]
  Contractual commitments addressing key requirements
\item[\(\checkmark\)]
  Third-party SOC 2 audit of security controls
\item[\(\checkmark\)]
  Legal counsel review confirming adequate SCCs
\item[\(\times\)]
  No independent technical audit of data handling practices
\item[\(\times\)]
  No verification mechanism for content deletion timeline
\item[\(\times\)]
  Limited visibility into sub-processor relationships beyond AWS
\end{itemize}

\textbf{Regulatory Risk Assessment:} Low to Medium

\begin{itemize}
\tightlist
\item
  \textbf{Current Status:} Compliant under accepted legal frameworks
  (SCCs remain valid post-\cite{schrems2020} with supplementary measures)
\item
  \textbf{Monitoring Required:} EU-US data transfer landscape
  evolves---new Data Privacy Framework, potential adequacy decisions, or
  regulatory guidance may affect compliance posture
\item
  \textbf{Mitigation in Place:} Mandatory user acknowledgment,
  transparent disclosure, alternative EU-hosted options available
\end{itemize}
\end{modelcardpanel}

\subsection{9. Costs and Resource
Usage}\label{9-costs-and-resource-usage}
\begin{modelcardpanel}
\textbf{Token Pricing (via Anthropic API):}

\begin{itemize}
\tightlist
\item
  Input tokens: \$15.00 per 1M tokens
\item
  Output tokens: \$75.00 per 1M tokens
\item
  Cached input tokens: \$1.50 per 1M tokens (90\% discount for repeated
  content)
\end{itemize}

\textbf{Typical Usage Cost Estimates:}

\begin{small}
\begin{tabular}{@{}lrrr@{}}
\textbf{Use Case} & \textbf{Input} & \textbf{Output} & \textbf{Cost} \\
\hline
Short query (question/answer) & 100 & 500 & €0.04 \\
Medium conversation (5-10 exchanges) & 1,000 & 2,000 & €0.17 \\
Extended analysis (document summary) & 20,000 & 10,000 & €1.05 \\
Research session (multi-turn reasoning) & 50,000 & 25,000 & €2.63 \\
Intensive coding project (architecture) & 100,000 & 50,000 & €5.25 \\
\end{tabular}
\end{small}

\smallskip
\noindent\textit{Token counts and costs are approximate. Input/Output measured in tokens.}

\textbf{Budget Impact for Fontys Platform:}

This model costs approximately \textbf{5x more than GPT-4o-mini}, our
baseline model for general use. That price premium buys:

\begin{itemize}
\tightlist
\item
  Superior reasoning for genuinely complex problems
\item
  Better context retention across long conversations
\item
  More nuanced understanding of ambiguous instructions
\item
  Stronger coding capabilities for sophisticated applications
\end{itemize}

\textbf{Cost-Justify Advanced Capabilities:} Before selecting this
model, ask yourself: "Does my task genuinely require these premium
capabilities, or would a cheaper EU-hosted alternative (GPT-4o via Azure
at €0.03-0.15 per 1M tokens) serve adequately?"

\textbf{Computational Intensity:} High

\begin{itemize}
\tightlist
\item
  Estimated energy per 1M tokens: \textasciitilde2-3 kWh (rough
  approx\-imation---Anthropic doesn\textquotesingle t publish exact
  figures)
\item
  Carbon footprint: Depends on AWS datacenter energy mix (partially
  renewable, exact percentage varies by region)
\item
  Environmental impact: Moderate to High relative to smaller models like
  GPT-4o-mini or Mistral Small
\end{itemize}

\textbf{Gateway Budget Controls:}

Our platform implements spending limits preventing runaway costs:

\begin{itemize}
\tightlist
\item
  \textbf{Per-conversation caps} preventing single session from
  consuming excessive budget
\item
  \textbf{Per-user monthly allocations} appropriate to role (higher for
  research, lower for coursework exploration)
\item
  \textbf{Per-model institutional limits} ensuring Claude Sonnet 4.5
  doesn\textquotesingle t dominate entire AI budget
\end{itemize}

\textbf{When approved for access, you\textquotesingle ll receive clear
budget allocation so you understand consumption limits before hitting
restrictions.}
\end{modelcardpanel}

\subsection{10. Sustainability Metrics}\label{10-sustainability-metrics}
\begin{modelcardpanel}
\textbf{Provider-Published Environmental Data:} Minimal

Anthropic does not publish model-specific carbon emissions, energy
consumption, or sustainability metrics. We lack visibility into:

\begin{itemize}
\tightlist
\item
  Exact kWh per million tokens processed
\item
  Carbon footprint per conversation
\item
  Energy source composition for inference datacenters
\item
  Water usage for cooling (relevant for large-scale GPU deployments)
\item
  E-waste from hardware lifecycle
\end{itemize}

\textbf{What We Can Estimate:}

Based on:

\begin{itemize}
\tightlist
\item
  Model architecture (large transformer, likely 100B+ parameters)
\item
  Inference hardware requirements (multiple high-end GPUs per request)
\item
  AWS datacenter energy profiles (mixed grid + renewable energy)
\end{itemize}

\textbf{Rough approximation:} Processing 1M tokens with Claude Sonnet
4.5 likely consumes 2-3 kWh, generating approximately 1-1.5 kg CO\textsubscript{2}
equivalent (varies significantly by AWS region energy mix).

\textbf{For context:} That\textquotesingle s roughly equivalent to:

\begin{itemize}
\tightlist
\item
  Driving an average car 5-7 km
\item
  Running a laptop continuously for 2-3 days
\item
  One load of laundry in an electric dryer
\end{itemize}

\textbf{Relative Environmental Impact:}

Compared to alternatives:

\begin{itemize}
\tightlist
\item
  \textbf{Higher impact than:} GPT-4o-mini, Mistral Small, Phi models
  (smaller/more efficient archi\-tectures)
\item
  \textbf{Similar impact to:} GPT-4o, Claude Opus, Mistral Large
  (comparable complexity models)
\item
  \textbf{Lower impact than:} Training models from scratch (inference is
  orders of magnitude less energy-intensive than training)
\end{itemize}

\textbf{AWS Sustainability Commitments:}

Anthropic\textquotesingle s infrastructure provider (AWS) has committed
to:

\begin{itemize}
\tightlist
\item
  100\% renewable energy by 2025 across global operations
\item
  Carbon neutrality for all operations
\item
  Water-positive goal for direct operations
\end{itemize}

\textbf{However:} These are AWS-wide commitments, not model-specific
guarantees. We cannot verify what percentage of Claude inference
actually runs on renewable energy versus fossil fuel-powered
datacenters.

\textbf{Institutional Environmental Consideration:}

\textbf{This represents a significant documentation gap.} Users
concerned about environmental impact deserve clear information enabling
sustainable AI choices. When environmental data remains unavailable, we
make this section explicitly acknowledge the absence rather than
omitting environmental considerations entirely.

\textbf{Climate-conscious users should:}

\begin{itemize}
\tightlist
\item
  Prefer smaller models (GPT-4o-mini, Mistral Small) when adequate for
  the task
\item
  Avoid unnecessary re-generation of content you already have
\item
  Batch queries rather than making many small requests
\item
  Consider whether AI tool use is necessary versus alternative
  approaches
\end{itemize}
\end{modelcardpanel}

\subsection{11. Comparable
Alternatives}\label{11-comparable-alternatives}
\begin{modelcardpanel}
\textbf{You don\textquotesingle t have to use this model.} Several
alternatives offer different trade-off profiles that may better suit
your needs:

\subsubsection{EU-Hosted Alternatives (Better Privacy, Lower
Cost)}\label{eu-hosted-alternatives-better-privacy-lower-cost}

\textbf{GPT-4o via Azure EU (Sweden/Germany datacenters)}

\begin{itemize}
\tightlist
\item
  \textbf{Processing location:} Guaranteed EU (Swedish or German
  datacenters)
\item
  \textbf{Capabilities:} Comparable reasoning, slightly faster
  inference, multimodal (can generate images)
\item
  \textbf{Cost:} €5-15 per 1M tokens (65-80\% cheaper)
\item
  \textbf{Trade-offs:} Slightly less nuanced instruction-following,
  context window 128K instead of 200K
\item
  \textbf{Best for:} Most use cases not requiring
  Anthropic\textquotesingle s specific reasoning style or extended
  context
\end{itemize}

\textbf{Mistral Large via Azure EU}

\begin{itemize}
\tightlist
\item
  \textbf{Processing location:} EU-hosted (Swedish datacenter)
\item
  \textbf{Capabilities:} Strong reasoning, efficient, European provider
\item
  \textbf{Cost:} €8-24 per 1M tokens (approximately 50\% cheaper)
\item
  \textbf{Trade-offs:} Smaller context window (128K), less established
  track record for complex coding
\item
  \textbf{Best for:} Users preferring European provider, general
  research applications, strong French language support
\end{itemize}

\subsubsection{Lower-Cost Alternatives (Same Geographic
Limitations)}\label{lower-cost-alternatives-same-geographic-limitations}

\textbf{Claude Haiku 4} (when available via our gateway)

\begin{itemize}
\tightlist
\item
  \textbf{Processing location:} US (same as Sonnet 4.5)
\item
  \textbf{Capabilities:} Faster, cheaper, adequate for straightforward
  tasks
\item
  \textbf{Cost:} \textasciitilde85\% cheaper than Sonnet 4.5
\item
  \textbf{Trade-offs:} Less sophisticated reasoning, shorter context
  window
\item
  \textbf{Best for:} Simple queries, quick code snippets, casual
  exploration
\end{itemize}

\textbf{GPT-4o-mini via Azure EU}

\begin{itemize}
\tightlist
\item
  \textbf{Processing location:} EU-hosted
\item
  \textbf{Capabilities:} Excellent for routine tasks, very fast
\item
  \textbf{Cost:} \textasciitilde90\% cheaper than Claude Sonnet 4.5
\item
  \textbf{Trade-offs:} Less capability for complex reasoning
\item
  \textbf{Best for:} Baseline use, practice, simple coursework support
\end{itemize}

\subsubsection{Maximum Capability
Alternative}\label{maximum-capability-alternative}

\textbf{Claude Opus 4.1} (restricted access)

\begin{itemize}
\tightlist
\item
  \textbf{Processing location:} US (same hosting limitations)
\item
  \textbf{Capabilities:} Anthropic\textquotesingle s most capable model,
  strongest reasoning
\item
  \textbf{Cost:} \textasciitilde2x more expensive than Sonnet 4.5
\item
  \textbf{Trade-offs:} Slower inference, significantly higher cost,
  restricted to specific research applications
\item
  \textbf{Best for:} Genuinely frontier research requiring absolute
  maximum capability
\end{itemize}

\textbf{Decision Framework:}

Ask yourself:

\begin{enumerate}
\def\labelenumi{\arabic{enumi}.}
\tightlist
\item
  \textbf{Does my task involve sensitive data?} → Prefer EU-hosted
  alternatives (GPT-4o, Mistral Large)
\item
  \textbf{Is extended reasoning truly necessary?} → If no, use
  GPT-4o-mini or Mistral Small
\item
  \textbf{Do I need Anthropic\textquotesingle s specific
  instruction-following style?} → If yes, Claude Sonnet 4.5 may be worth
  premium
\item
  \textbf{Is cost a significant concern?} → EU alternatives provide
  50-80\% savings with comparable capability
\end{enumerate}

\textbf{Our Platform Makes Switching Easy:} All these alternatives are
available through the same interface. Try cheaper EU-hosted models
first---you can always escalate to Claude Sonnet 4.5 if genuinely
needed.
\end{modelcardpanel}

\subsection{12. Governance Status and Institutional
Oversight}\label{12-governance-status-and-institutional-oversight}
\begin{modelcardpanel}
\textbf{Current Access Configuration:}

\(\checkmark\) \textbf{Automatically Available To:}

\begin{itemize}
\tightlist
\item
  MA-AAI1 Advanced students (with faculty authorization)
\item
  Faculty with research authorization via professorship groups
\item
  Administrative staff with documented need
\end{itemize}

\textbf{Warning}\,\textbf{:} \textbf{Requires Access Request:}

\begin{itemize}
\tightlist
\item
  Undergraduate students seeking advanced capabilities
\item
  Research projects involving high-intensity usage
\item
  Any user outside pre-authorized Active Directory groups
\end{itemize}

\(\times\) \textbf{Not Currently Available:}

\begin{itemize}
\tightlist
\item
  Guest accounts
\item
  External collaborators (unless specifically provisioned)
\end{itemize}

\textbf{Access Rationale:}

Claude Sonnet 4.5 receives restricted deployment because:

\begin{enumerate}
\def\labelenumi{\arabic{enumi}.}
\tightlist
\item
  \textbf{US hosting creates privacy considerations} requiring user
  understanding of international data transfer implications
\item
  \textbf{High costs demand justification} ensuring institutional
  resources distribute appropriately rather than concentrating in
  premium services
\item
  \textbf{Powerful capabilities warrant user sophistication} for
  appropriate use and critical output evaluation
\item
  \textbf{Training data opacity requires informed consent} acknowledging
  evaluation limitations
\end{enumerate}

We do NOT restrict access as arbitrary gatekeeping. Restrictions ensure
users:

\begin{itemize}
\tightlist
\item
  Understand privacy trade-offs they\textquotesingle re accepting
\item
  Have legitimate needs genuinely requiring premium capabilities
\item
  Won\textquotesingle t inadvertently consume disproportionate
  institutional budget without awareness
\end{itemize}

\textbf{Access Request Requirements:}

To request access, you must provide:

\begin{enumerate}
\def\labelenumi{\arabic{enumi}.}
\tightlist
\item
  \textbf{Use Case Description} substantiating need for capabilities not
  satisfied by EU-hosted alternatives like GPT-4o or Mistral Large (be
  specific---"I need advanced reasoning" isn\textquotesingle t
  sufficient; explain what complex problem requires this particular
  model)
\item
  \textbf{Privacy Acknowledgment} demonstrating understanding that your
  prompts process in US data\-centers under different privacy frameworks
  than EU alternatives
\item
  \textbf{Faculty Authorization} (for students) confirming course
  requirements genuinely necessitate this specific
  model\textquotesingle s capabilities
\item
  \textbf{Budget Justification} (for high-intensity research) explaining
  expected usage patterns and why EU alternatives won\textquotesingle t
  serve your application
\end{enumerate}

\textbf{Requests typically receive response within 3-5 business days.}
We approve most requests demonstrating substantiated need---the process
exists to ensure informed choice, not to deny access arbitrarily.

\textbf{Required Acknowledgments Before First Use:}

Even after access approval, you must explicitly acknowledge:

\begin{itemize}
\tightlist
\item[$\square$]
  My data will process in United States datacenters subject to US legal
  jurisdiction
\item[$\square$]
  Anthropic has not disclosed training data sources; Fontys cannot
  verify bias mitigation or content appropriateness
\item[$\square$]
  This model may generate confident but incorrect information; I will
  verify critical facts independently
\item[$\square$]
  Cost implications understood---this model costs 5x baseline
  alternatives, and I\textquotesingle ve considered whether EU-hosted
  options would meet my needs
\end{itemize}

\textbf{Harm Reporting Procedures:}

If you encounter problems, contact:

\begin{itemize}
\tightlist
\item
  \textbf{Problematic outputs} (bias, harmful content, persistent
  inaccuracy):
  \href{mailto:ai-governance@fontys.nl}{\href{mailto:ai-governance@fontys.nl}{\nolinkurl{ai-governance@fontys.nl}}}
\item
  \textbf{Privacy concerns} (unexpected data handling, access issues):
  \href{mailto:privacy-office@fontys.nl}{\href{mailto:privacy-office@fontys.nl}{\nolinkurl{privacy-office@fontys.nl}}}
\item
  \textbf{Technical failures} (model unavailable, gateway errors):
  \href{mailto:ict-support@fontys.nl}{\href{mailto:ict-support@fontys.nl}{\nolinkurl{ict-support@fontys.nl}}}
\item
  \textbf{Budget/access questions}:
  \href{mailto:platform-admin@fontys.nl}{\href{mailto:platform-admin@fontys.nl}{\nolinkurl{platform-admin@fontys.nl}}}
\end{itemize}

\textbf{Liability Clarification:}

\begin{itemize}
\tightlist
\item
  \textbf{Provider:} Anthropic\textquotesingle s terms limit liability
  for model outputs and do not guarantee accuracy or suitability for any
  purpose
\item
  \textbf{Institution:} Fontys provides access through our gateway but
  cannot verify all provider claims; we document known limitations and
  gaps honestly
\item
  \textbf{You:} Responsible for validating outputs before relying on
  them for decisions, coursework, or research. Human oversight required
  for high-stakes applications.
\end{itemize}

\textbf{This model is a tool, not an oracle.} Treat its outputs as
sophisticated but fallible suggestions requiring your judgment and
verification.
\end{modelcardpanel}

\subsection{Summary: Should You Use Claude Sonnet
4.5?}\label{summary-should-you-use-claude-sonnet-45}
\begin{modelcardpanel}
\textbf{Use this model when:}

\begin{itemize}
\tightlist
\item
  Your task genuinely requires extended reasoning across complex,
  ambiguous problems
\item
  You\textquotesingle re working with very long documents (100+ pages)
  requiring comprehensive analysis
\item
  You need sophisticated coding support for architectural decisions, not
  just syntax
\item
  EU-hosted alternatives have proven inadequate for your specific
  application
\item
  You understand and accept US data processing implications
\end{itemize}

\textbf{Consider alternatives when:}

\begin{itemize}
\tightlist
\item
  Your task is straightforward or routine (use GPT-4o-mini---same
  quality, 5x cheaper)
\item
  Data sensitivity requires EU-only processing (use GPT-4o or Mistral
  Large via Azure EU)
\item
  You\textquotesingle re exploring casually or practicing (reserve
  expensive models for genuine need)
\item
  Cost efficiency matters and cheaper options might suffice
\end{itemize}

\textbf{Remember:}

\begin{itemize}
\tightlist
\item
  This model excels at genuine complexity but costs premium---use
  judiciously
\item
  Geographic limitations create privacy trade-offs requiring informed
  consent
\item
  Training data opacity means we cannot verify all provider claims
\item
  Alternative models offer comparable capability with better
  privacy/cost profiles
\end{itemize}

\textbf{Need access?} Submit request via platform interface with
substantiated use case.

\textbf{Have questions?} Contact
\href{mailto:ai-governance@fontys.nl}{\href{mailto:ai-governance@fontys.nl}{\nolinkurl{ai-governance@fontys.nl}}}
\end{modelcardpanel}
\end{tcolorbox}

\clearpage
\bibliographystyle{apacite}
\bibliography{whitepaper}

\end{document}